\documentclass[10pt,preprint]{aastex}

\def \etal   {\hbox{\it et~al.\/}}
\def \kms    {km~s$^{-1}$}

\def\half{\hbox{$\frac{1}{2}$}}

\def\ie{\hbox{i.e.,}}

\newcommand{\vinf}{\mbox{$v_{\infty}$}}
\newcommand{\mui}{\mbox{$\cos \theta_{\rm i}$}}
\newcommand{\mus}{\mbox{$\cos \theta_{\rm s}$}}
\newcommand{\xxi}{\mbox{$\sin \theta_{\rm i}$}}
\newcommand{\xxs}{\mbox{$\sin \theta_{\rm s}$}}
\newcommand{\muisq}{\mbox{$\cos^2 \theta_{\rm i}$}}
\newcommand{\mussq}{\mbox{$\cos^2 \theta_{\rm s}$}}
\newcommand{\xxisq}{\mbox{$\sin^2 \theta_{\rm i}$}}
\newcommand{\xxssq}{\mbox{$\sin^2 \theta_{\rm s}$}}

\begin{document}

%
%

\title{The Hanle Effect as a Diagnostic of Magnetic Fields in Stellar
Envelopes IV.  Application to Polarized P~Cygni Wind Lines}

\author{Richard\ Ignace \footnote{Currently at East Tennessee State University; Email:  ignace@etsu.edu},
        Kenneth\ H.\ Nordsieck, and
        Joseph\ P.\ Cassinelli }
\affil{
        Department of Astronomy,
        University of Wisconsin,
        5534 Sterling Hall,
        475 N.\ Charter St.,
        Madison, WI  53706-1582 }


\begin{abstract}

The Hanle effect has been proposed as a new diagnostic of
circumstellar magnetic fields for early-type stars, for which
it is sensitive to field strengths in the 1-300~G range. In
this paper we compute the polarized P-Cygni line profiles that
result from the Hanle effect. For modeling the polarization,
we employ a variant of the ``last scattering approximation''.
For cases in which the Sobolev optical depths are greater than
unity, the emergent line intensity is assumed to be unpolarized;
while for smaller optical depths, the Stokes source functions
for the Hanle effect with optically thin line scattering are
used. For a typical P~Cygni line, the polarized emission forms
in the outer wind, because the Sobolev optical depth is large
at the inner wind.  For low surface field strengths, weak
P~Cygni lines are needed to measure the circumstellar field.
For high values of the surface fields, both the Zeeman and Hanle
diagnostics can be used, with the Zeeman effect probing the
photospheric magnetic fields, and the Hanle effect measuring
the magnetic field in the wind flow.  Polarized line profiles
are calculated for a self-consistent structure of the flow
and the magnetic geometry based on the WCFields model, which
is applicable to slowly rotating stellar winds with magnetic
fields drawn out by the gas flow. For surface fields of a few
hundred Gauss, we find that the Hanle effect can produce line
polarizations in the range of a few tenths of a percent up to
about 2 percent.

\end{abstract}

\section{INTRODUCTION}

The Hanle effect is a diagnostic that refers to the modification of
resonance line scattering polarization in the presence of a magnetic
field. The effect begins to have an influence on the polarization  at
fairly small field strengths of just a few Gauss with sensitivity to
fields up to around 300~Gauss.  Experiments to describe the polarization
from resonance line scattering date back primarily back to the first
third of the 20th century (see Mitchell \& Zemansky 1934).  The influence
of a magnetic field on the line polarization was explained first by
a young physicist named Wilhelm Hanle.  Hanle (1924) described the
change of linear polarization by the magnetic field in semi-classical
terms as arising from the precession of an atomic, damped, harmonic
oscillator. From a quantum mechanical point-of-view, the effect is
understood in terms of interferences that occur when the degeneracy of
the magnetic sublevels in the excited state is partially lifted. The
effect has come to have applications to many topics in atomic physics
as described in Moruzzi \& Strumia (1991).

Astrophysically, the Hanle effect has been used only in the Sun.
Applications include magnetic field measurements in the chromosphere,
corona, and particularly in prominences and filaments (e.g., see Lin,
Penn, \& Kuhn 1998).  A detailed description of the physics of the Hanle
effect and polarized radiation transport for solar studies appears in
Stenflo (1994).  However, for stellar astronomy, the Hanle effect remains
relatively unknown.

This paper is the fourth in a series to explore applications of the effect
for other stars, especially hot stars with winds (Ignace, Nordsieck,
\& Cassinelli 1997 [Paper~I]; Ignace, Cassinelli, \& Nordsieck 1999
[Paper~II]; Ignace 2001 [Paper~III]). It has been our goal to develop
model line profiles with the Hanle effect that can be used to interpret
spectropolarimetric data for inferring the properties of circumstellar
magnetic fields.  The Zeeman effect has had some success in measuring
{\it photospheric} surface fields for a few hot stars that are not part
of the extreme Bp class (e.g., Henrichs 2003; Donati \etal\ 2001, 2002;
and Neiner 2002). However, Ignace \& Gayley (2003) have shown that the
Zeeman effect suffers serious technical challenges for use in diagnosing
circumstellar magnetic fields from wind emission lines.  For example,
typical circular polarizations for such lines have peak values of only
0.01\% for a field with a surface value of around 100~G. Fields of this
magnitude can have a significant influence on the wind flow from hot
stars (e.g., Maheswaran \& Cassinelli 1992; Babel \& Montmerle 1997ab;
Cassinelli \etal\ 2002; ud-Doula \& Owocki 2002).  Small polarizations
result for the Zeeman effect because the Zeeman-split components that
produce the circular polarization are incoherent and oppositely signed.
The wavelength separation of the Zeeman split lines is small compared
with typical astrophysical broadening effects.  Consequently, the
net circular polarization of the line is severely reduced owing to
polarimetric cancellation from blending of the Zeeman components.
The Hanle effect is a {\em modification} of the linear polarization
that arises from resonance line scattering that does not suffer from
polarimetric cancellation due to line blending.

\begin{figure}
\plotone{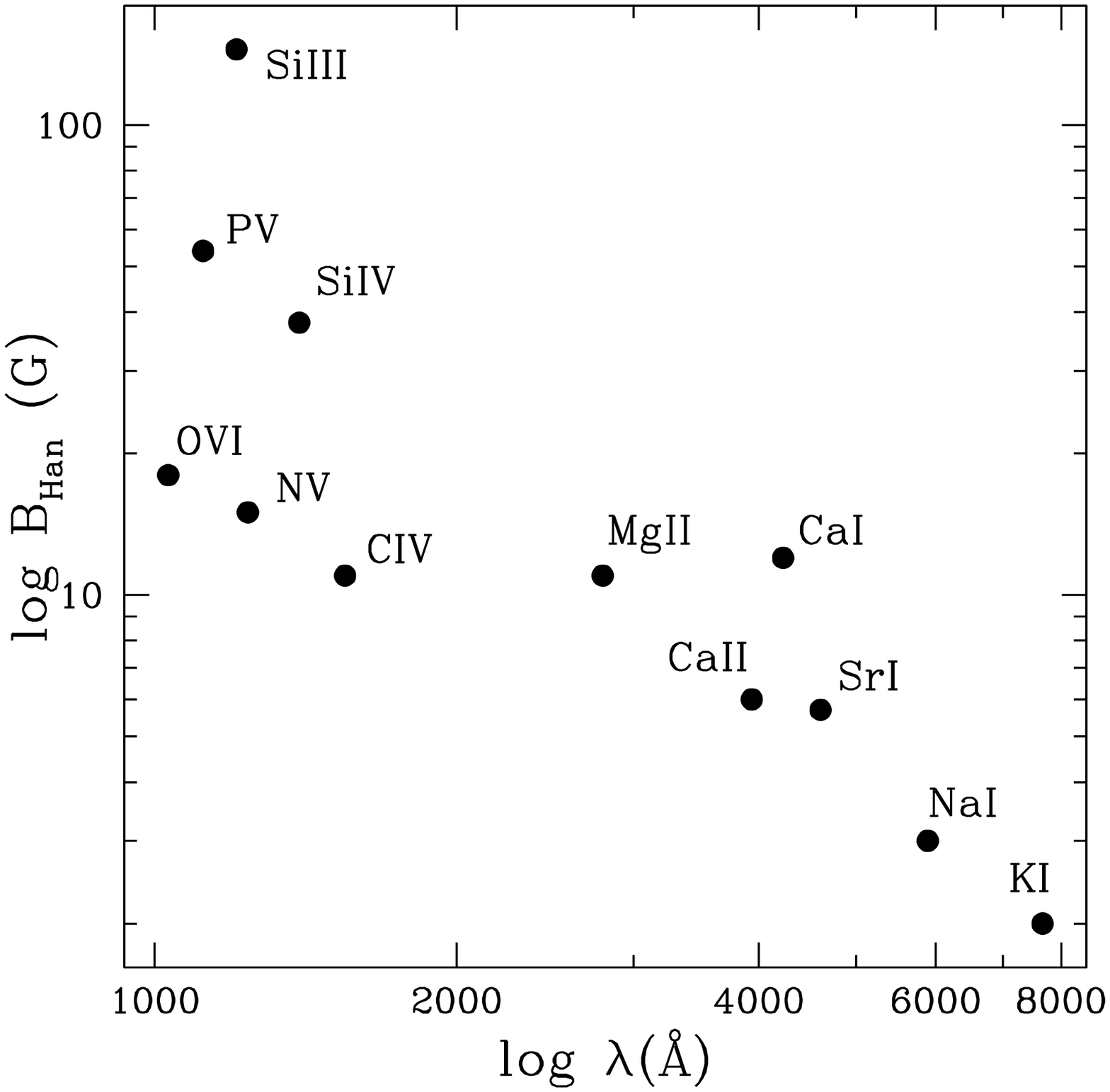}
\caption[]{
A plot of the Hanle field strength $B_{\rm Han}$ for resonance
scattering lines common to astrophysics.  Note that a few of
these lines are singlets, but most are Li-like doublets.  Of
the doublets, only the shorter wavelength component will show
a Hanle effect, since the longer wavelength component scatters
isotropically and produces no line polarization.  Evident is
that shorter wavelength lines, having higher $A$-values, tend
to have higher Hanle field values.  The shorter wavelength
lines are also associated with more highly ionized atoms that are
commonly observed in hot star winds.

\label{fig1}}

\end{figure}

In a semi-classical description of the Hanle effect, the line scattering
can be split into two parts: an isotropic scattering component and a
dipole scattering component like free electrons (e.g., Hamilton 1947;
Chandrasekhar 1960).  The fraction of the scattering that is dipole-like
is referred to as the ``polarizability'', $E_1$; the isotropic scattering
contribution is given fractionally as $(1-E_1)$.  The introduction
of a magnetic field induces a precession of the classical oscillator
motion of the atomic scatterers.  At the same time, the oscillations are
damped, emitting their absorbed radiation at a rate given by the Einstein
$A$-value for the transition of interest.  Thus there is a competition
between the radiative rate versus the Larmor precessional rate that has
the frequency, $\omega_L = e B /m_e c$.  Using these same arguments,
Hanle (1924) deduced that a significant effect results when these two
rates are somewhat comparable.  We thus define a ``Hanle ratio'' given by

\begin{equation}
\frac{B}{B_{\rm Han}} = \frac{2g_L \omega_L}{A_{ul}}
	= g_L\,\frac{(B/5~G)}{(A_{ul}/10^8~s^{-1})},
\end{equation}

\noindent where $g_L$ is the Lande factor, and $A_{ul}$ is the Einstein
$A$-value from upper level $u$ to lower level $l$.  The Hanle field strength
$B_{\rm Han}$ is a conveniently defined parameter given by 

\begin{equation}
B_{\rm Han} = \frac{m_e\,c\,A_{ul}}{2\,g_L\,e},
\end{equation}

\noindent that characterizes the magnetic field scale at which the the
Hanle effect is important.  Values of $B_{\rm Han}$ for typical strong
resonance lines common to astrophysics are displayed in Figure~\ref{fig1}.
Note that the lines at shorter wavelengths tend to have larger Hanle
fields, because $A$-values increase roughly as $\lambda^{-2}$.

Recognizing that the Hanle effect is sensitive to modest field values of
10--100~G, we explored in Paper~I its use for determining stellar wind
magnetic field values and geometries from the total line polarization of
optically thin emission lines. In Paper~II we considered the Hanle effect
in optically thin and resolved line profiles that from in equatorial
disks.  In Paper~III expressions were determined for the depolarizing
effect of having stellar radiation arising from a finite disk instead of
from a point star. For thin lines the correction was found to be the same
as that for electron scattering polarization as derived by Cassinelli,
Nordsieck, \& Murison (1987). It was pointed out by Cassinelli, Nordsieck,
\& Ignace (2001) that in contrast with the solar case, the application of
the Hanle effect to hot stars is somewhat simpler and less ambiguous. This
is for two reasons. (a) The hot star wind lines are highly NLTE so that
virtually all bound electrons of metal species are in the ground state,
whereas detailed NLTE considerations are required in solar applications,
and (b) the stellar continuum radiation that is being scattered is
relatively ``flat'' over the frequency width of any given line, while
solar applications often involve scattering of chromospheric emission
lines for which the flux changes strongly with wavelength
across the resonance line profiles.

The goal of these papers has been to move from an introduction of the Hanle
effect for the broader stellar community to an investigation of its use by
way of model calculations of increasing physical realism and numerical
sophistication. In this paper we treat more realistic and more dynamically
self-consistent magnetic field distributions in the wind, and we present
results for the distribution of polarization across P~Cygni line
profiles.  

Section~2 outlines the line polarization calculations, and introduces a
``single scattering'' approximation that is used together with traditional
Sobolev line profile analyses.  We find that line optical depth effects
have a significant influence on the polarized line profiles.  Heuristic
results are described in \S 3, including simple spherical shells with
axial fields, and spherical winds with dipole fields.  Results for a more
realistic scenario for the field and flow are presented \S 4. The model
chosen for this is the slow magnetic rotator field configuration based on
the WCFields theory by Ignace, Cassinelli, \& Bjorkman (1998). Significant
line polarizations of around 1\% are found to result.  Lastly, \S 5
offers a discussion of our modeling, with implications for observations
of hot star winds at ultraviolet wavelengths and future directions for
modeling efforts.

\section{MODEL LINE PROFILES}

In the previous papers, we made calculations of the Hanle effect in emission
lines that are valid only in the optically thin case.  Here we extend the
results to allow for optical depth effects, by implementing escape
probability theory and introducing the ``single scattering approximation''
for the emergent polarization from a volume sector. In this approximation
the radiation at large optical depth is assumed completely unpolarized,
and magnetically sensitive line polarization effects arise only in
optically thin regions where a photon is considered to scatter only once.
We make use of the Sobolev theory for resonance line scattering in stellar
winds as developed by Sobolev (1960), Castor (1970), and Mihalas (1978).
In Sobolev theory the radiative transfer is approximated as being local,
because of the fast flow speeds and large velocity gradients that are
present in stellar winds. The scattering and production of photons at
any small volume in the medium can be assumed to have no interaction
with any other portion of the flow.

\subsection{Standard Sobolev Theory}

In the standard Sobolev approach (e.g., as summarized in
Mihalas 1978), the observable emission line
profile is constructed from a ray-by-ray determination of the emergent line
intensity for a fixed frequency within the line
profile.  An integration of these intensities is used to obtain the total
emission or absorption at a particular frequency within the profile.
The key variables for computing the line emission for a spherical flow are
the velocity shift along the line-of-sight axis $z$,

\begin{equation}
v_{\rm z } = - v(r)\,\mu ,
	\label{eq:velshift}
\end{equation}

\noindent where $v$ is the spherical velocity flow, $r$ is the
radial distance from the star, and $\mu=\cos \chi$ is the polar spherical
angle from the observer's line-of-sight; the Sobolev optical depth

\begin{equation}
\tau_S = \frac{\kappa_l \,\rho \,\lambda_0}{(v/r)\,(1+\sigma\,\mu^2)},
	\label{eq:tsob}
\end{equation}

\noindent where $\kappa_l$ is the frequency-integrated line opacity,
$\rho=\rho(r)$ is the density, $\lambda_0$ is the line central wavelength,
and $\sigma$ is related to the line-of-sight velocity gradient (as will
be given in eq.~[\ref{eq:sigma}]); and the emergent intensity along the
ray at impact parameter $p$,

\begin{equation}
I_\nu(p) = S_\nu(r)\,\left(1-e^{-\tau_S}\right),
	\label{eq:Iemerge}
\end{equation}

\noindent where $S_\nu (r)$ is the line source function.

Because the flow speed is much greater than the thermal broadening
only a small distance along the line-of-sight affects the intensity,
and the integration over impact parameters at a fixed frequency in the
line profile corresponds to an integral over  an ``isovelocity surface''.
Such a surface is defined by  $v_{\rm z} = constant$.  A simplification
implicit in equation~(\ref{eq:Iemerge}) is that the surfaces of constant
velocity-shift are single-valued with respect to intercepting rays
of fixed $p$. More complicated cases involving non-local coupling
associated with doublets or more general flow patterns are presented
by Rybicki \& Hummer (1978).  Also common in applications of Sobolev
theory is the distinct ``core-plus-halo'' approximation in which there
is a continuum-producing photosphere and a distinct overlying outflow,
in which the absorption of the continuum is computed via $I_* e^{-\tau_S}$
along each ray intercepting the photosphere.

A different but equivalent approach based on escape probabilities was
developed by Rybicki \& Hummer (1983).  The method involves integrating
the emissivity for the line radiation over the emitting volume accounting
for optical depth effects via escape probabilities.  This approach
is more suitable for our purposes not because of non-local coupling
or complicated flow patterns (for we assume a radial velocity flow),
but because we want to allow for axisymmetric magnetic field topologies
viewed from arbitrary viewing inclinations.  It is convenient to solve
for the emergent polarized flux based on a volume integration over the
scattering medium.

To describe our calculations, we begin with the basic escape probability
approach to Sobolev theory.  From Rybicki \& Hummer (1983), the emergent
luminosity from a volume element is

\begin{equation}
dL_\nu =  4\pi\, j_\nu\,\rho\,p_\mu\,r^2\, dr\, d\alpha\,d\mu,
\end{equation}

\noindent where $j_\nu\rho$ is the emissivity as given by

\begin{equation}
j_\nu\,\rho = \kappa_l \, \rho \, S_\nu \, \psi(\nu-\nu_{\rm z})
	\label{eq:jnu}
\end{equation}

\noindent and $p_\mu$ is the directional escape probability given by

\begin{equation}
p_\mu = \frac{1-e^{-\tau_S}}{\tau_S}
	\label{Eq:pesc}
\end{equation}

\noindent for $\tau_S = \tau_S(r,\mu)$ the Sobolev optical depth as
given previously.  In equation~(\ref{eq:jnu}), the factor $\psi$ is the
line profile function.  In Sobolev theory this function is traditionally
a delta function; however, it is key only that this function be much
more narrow than the wind-broadening of the line.  For the fast winds
of hot stars, this turns out to be an excellent approximation, since
in the co-moving frame of the gas, $\psi$ has a velocity width of
order $v_{\rm th}$ for a given ion species, and this
is much smaller than the wind broadening that is of order $v_\infty$.
The frequency $\nu_{\rm z}$ appearing in the argument of $\psi$ is
related to the velocity shift of equation~(\ref{eq:velshift}) by

\begin{equation}
\nu_{\rm z} = \nu_0\,\left(1-\frac{v_{\rm z}}{c}\right),
\end{equation}

\noindent for $\nu_0$ the central wavelength of the line under
consideration.  The narrow profile response function $\psi$ effectively
picks out, for a given sightline, a particular volume element or ``cell''
corresponding to that cell's Doppler shift with $\nu=\nu_{\rm z}$.
Hence the strong velocity gradients in hot star winds ensures that the
radiative transfer for all volume elements is ``local'', meaning all
the cells are radiatively de-coupled from each other.

The profile function $\psi$ is defined such that

\begin{equation}
\int_0^{\infty}\,\psi(\nu-\nu') \, d\nu = 1.
\end{equation}

\noindent For the purposes of numerical calculation, we consider a
wind emission line profile to have a number of fixed frequency bins, each
of width $\Delta \nu_{\rm z} = (\nu_0/c)\,\Delta v_{\rm z} = \Delta v_{\rm
z}/\lambda_0$ and in total spanning the full line width $\Delta \nu_\infty
= (2v_\infty/c)\,\nu_0$.  Then for a given volumetric cell existing in a
spherical shell at fixed radius $r$, with fixed radial extent $\Delta r$,
the luminosity contribution to the observed line emission will be

\begin{equation}
\Delta L_\nu = 4\pi\,\kappa_l \, \rho \, S_\nu \,p_\mu\,r^2\,\Delta r\,\Delta
	\alpha\,\frac{\Delta \mu}{\Delta \nu_{\rm z}}.
\end{equation}

\noindent Using the expression for the escape probability
equation~(\ref{Eq:pesc}) and the Sobolev optical depth
equation~(\ref{eq:tsob}), and the relation between $\Delta \nu_{\rm z}$
and $\Delta v_{\rm z}$, the luminosity from this cell becomes

\begin{equation}
\Delta L_\nu = 4\pi\,S_\nu\,\left(1-e^{-\tau_S}\right)\,
	(1+\sigma\,\mu^2)\,v(r)\,
	r\,\Delta r\,\Delta \alpha\, \frac{\Delta \mu}{\Delta v_{\rm z}},
\label{eq:delLnu}
\end{equation}

\noindent where

\begin{equation}
\sigma = \frac{d\ln v}{d\ln r} - 1.
	\label{eq:sigma}
\end{equation}

\noindent It is the form of  the Sobolev approximation given in
equation~(\ref{eq:delLnu}) that we use for computing emission line
profiles.  In particular, pure scattering resonance lines are assumed,
for which the source function is 

\begin{equation}
S_\nu = \frac{\beta_{\rm c}}{\beta_{\rm esc}}\,I_*,
	\label{eq:Snu}
\end{equation}

\noindent where $\beta$ represents the escape probability averaged over the
relevant solid angle,

\begin{equation}
\beta (\mu) = \frac{1}{4\pi}\,\int_\Omega \, p_\mu \, d\Omega
	= \frac{1}{2}\,\int_\mu^{+1}\, p_\mu\,d\mu.
\end{equation}

\noindent In particular, the two $\beta$'s that occur in
equation~(\ref{eq:Snu}) are: the penetration factor

\begin{equation}
\beta_{\rm c} = \frac{1}{2}\,\int_{\mu_{\rm c}}^{+1}\, p_\mu\,d\mu,
\end{equation}

\noindent for $\mu_{\rm c} = \sqrt{1-R_*^2/r^2}$, and the escape factor

\begin{equation}
\beta_{\rm esc} = \frac{1}{2}\,\int_{-1}^{+1}\, p_\mu\,d\mu.
\end{equation}

It is useful to compare the thin and thick limits for the source
function.  For regions where $\tau_S \ll 1$ for all $\mu$, one has from
equation~(\ref{Eq:pesc})
that $p_\mu \approx 1 - {\cal O}(\tau_S)$.  In this case $\beta_{\rm c}
\approx 0.5 (1-\mu_{\rm c})$ and $\beta_{\rm esc} \approx 1$. Thus the ratio
of $\beta$'s yields the well-known dilution factor 

\begin{equation}
W(r) = \frac{1}{2}\,\left( 1-\sqrt{1-\frac{R_*^2}{r^2}}\right).
\end{equation}

The thick limit with $\tau_S \gg 1$ is quite different. The
escape probability becomes $p_\mu \approx \tau_S^{-1} =
f(r)\,[1+\sigma(r)\mu^2]$, for $f(r)$ a function of radius only. For
a spherical shell of radius $r$, this escape probability factor is
a parabola in $\mu$, and thus a parabola also in Doppler shift in
contributing to the line profile emission.  For typical wind velocity laws
that increase monotonically from some small value to an asymptotically
large value, $\sigma$ will range from $\sigma \approx -1$ to $\sigma
\gg 1$.  This implies that the parabolic factor can be either concave up
or down, and this behavior affecs the line profile. The exact
form of the source function thus depends on the velocity law $v(r)$, and for
our model line profile calculations, a typical beta wind velocity law,
which to avoid confusion with the escape probabilities) will be written
using $\gamma$, is adopted as given by

\begin{equation}
v(r) = v_\infty\,\left(1-\frac{bR_*}{r}\right)^\gamma,
\end{equation}

\noindent where $\gamma$ is the velocity law exponent,
and $b < 1$ sets the initial wind speed $v_0$.  For such a form, and with
$\gamma =1$, it can be shown that in the optically thick limit, the source
function will vary as $S_\nu \propto r^{-3}$.

Having described the source functions for wind emission lines in the
standard Sobolev approach, we turn now to a discussion of the polarizing
properties of the Hanle effect.

\subsection{The Hanle Source Functions for Single Scattering}

\begin{figure}
\plotone{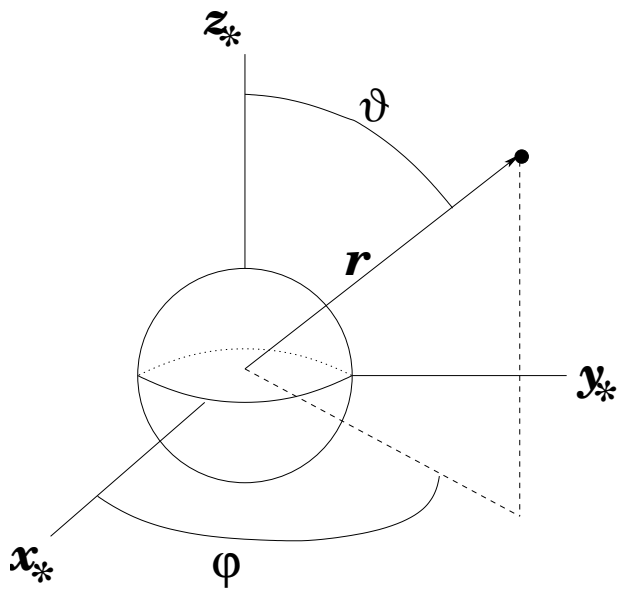}
\caption[]{
The stellar coordinate system, both Cartesian and spherical.  The
$Z_*$-axis is taken as the reference direction for the field geometry.
If the direction $\hat{r}$ corresponds to the illumination ray of a
scattering point at the dot, then $\vartheta=\vartheta_{\rm i}$ and
$\varphi=\varphi_{\rm i}$.

\label{fig2}}
\end{figure}

\begin{figure}
\plotone{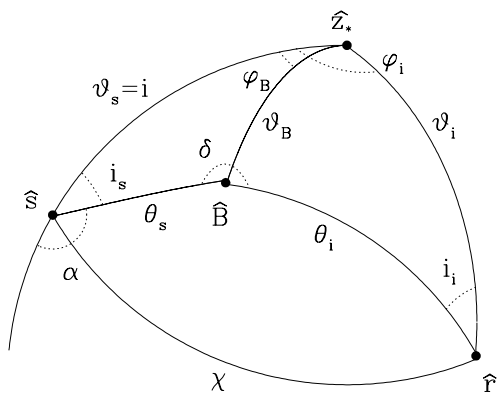}
\caption[]{
The scattering geometry centered at the scattering point of
Fig.~\ref{fig2}.  The unit vectors $\hat{r}$ and $\hat{s}$ are for
the radial direction and the scattering direction.  The angles $(\chi,
\alpha)$ are polar spherical angles for the observer (along $\hat{s}$).
The angles $\vartheta_B$ and $\varphi_B$ specify the vector orientation
of the local magnetic field, $B$.  The angles $(\theta, \phi)$ are polar
spherical angles for a system defined by this local magnetic field.
The angle $i_{\rm s}$ is a Mueller rotation angle that rotates the $Q$
and $U$ polarizations from the reference frame of the magnetic field to
that of the observer's measurement axes.  \label{fig3}}

\end{figure}

Computing the polarimetric properties of a beam of line radiation
that is modified by the Hanle effect requires a number of coordinate
transformations involving the local magnetic field at the scattering
point, the star-centered frame, and the observer's coordinate frame.
Figure~\ref{fig2} shows the stellar coordinates that are Cartesian $(x_*,
y_*, z_*)$ and spherical $(r, \vartheta, \varphi)$.  The solid point in
Figure~\ref{fig2} represents a point of scattering, with the scattering
geometry shown in Figure~\ref{fig3}. There, the unit vector $\hat{s}$
for the scattered light is the direction toward the observer (\ie\ along
the $z$-axis).  The spherical angle $\chi$ is the angle of scattering
from the radial direction, and $\alpha$ is an angle to the local field
direction. Cylindrical coordinates $(p, \alpha, z)$ refer to the observer,
with $p$ being the impact parameter (not shown in the figure).  We take
the star to be axisymmetric, and tilted with respect to the observer's
axis by an inclination angle $i$.

The field topology is also assumed to be axisymmetric about the axis
$z_*$, and star-centered.  In the star system, the orientation of the
vector magnetic field at any point in the wind is given by the polar
angles $\vartheta_B$ and $\varphi_B$.  Polar angles $\theta_{\rm s}$ and
$\theta_{\rm i}$ refer to the direction, measured from the local magnetic
field axis $\hat{B}$, of the scattered light and the radial direction,
where the local radius vector $\hat{r}$ is taken to be the symmetry
axis for the incident bundle of intensities from the stellar atmosphere.
We will ignore limb darkening so that stellar intensities $I_*$ are the
same from every point of the projected photosphere.  Azimuthal angles
$\phi_{\rm s}$ and $\phi_{\rm i}$ are measured around $\hat{B}$.  It is
only their difference that enters into the phase function that governs
the Hanle effect, so we define $\delta = \phi_{\rm s} - \phi_{\rm i}$,
which is what appears in Figure~\ref{fig3}.  Details regarding the
inter-relations of the various geometrical angles and their evaluation
are given in Appendix~\ref{appA}.

The Stokes parameters $I$, $Q$, $U$, and $V$ are used to describe the
polarization of the light, which will be represented as a ``vector'',
as for example the flux of light will be ${\bf F} = (F_I, F_Q, F_U,
F_V)={\bf L}/4\pi D^2$.  However, we shall always assume that $V=0$
(i.e., the circular polarization is zero), and so will ignore that component
in what follows.

For the Stokes source function, we have ${\bf S} = (S_I, S_Q, S_U)$.
In Papers I--III, expressions were given for the source function elements,
which we are repeated here in modified form:

\begin{eqnarray}
S_I & = & \;\;S_{II} , \\
S_Q & = & \;\;S_{QI} \cos 2i_{\rm s} + S_{UI} \sin 2i_{\rm s} , \\
S_U & = & -S_{QI} \sin 2i_{\rm s} + S_{UI}  \cos 2i_{\rm s},
\end{eqnarray}

\noindent where $S_{II}$, $S_{QI}$, and $S_{UI}$ are the source
functions in the magnetic field system, and $i_{\rm s}$ is an azimuthal 
angle between that system and the observer's (see Fig.~\ref{fig2}).
The source function components are,

\begin{eqnarray}
S_{II} & = & J + \frac{3}{8}\,E_1\,(3K-J)\,\left[ \frac{1}{6}(1-3
	\muisq)\,(1-3\mussq) + a(B,\delta)\,\mus\,\mui\,\xxs\,\xxi \right. \nonumber \\
 &   &	\left. + \frac{1}{2} c(B,\delta)\,\muisq\,\xxssq \right] , \label{eq:SII} \\
S_{QI} & = & \frac{3}{8} \,E_1\,(3K-J)\,\left[ \frac{1}{2}\xxssq\,
	(1-3\muisq) \right. \nonumber \\
 &   &	\left. + a(B,\delta)\,\mus\,\mui\,\xxs\,\xxi - \frac{1}{2}c(B,\delta)
	\,\xxisq\,(1-\mussq) \right], \\
S_{UI} & = & \frac{3}{8} \,E_1\,(3K-J)\,\left[ -b(B,\delta)\,\mui\,\xxi\,\xxs
	+ d(B,\delta)\,\mus\,\xxisq \right],	\label{eq:SUI}
\end{eqnarray}

\noindent where $E_1$ is the polarizability, $J$ and $K$ are the standard
Eddington moments of the radiation field (assumed to be determined by
the incident stellar radiation field), and the local field strength
enters through the functions $a$, $b$, $c$, and $d$, which are given by

\begin{eqnarray}
a & = & \cos \alpha_1 \cos (\delta-\alpha_1), \\
b & = & \cos \alpha_1 \sin (\delta-\alpha_1), \\
c & = & \cos \alpha_2 \cos 2(\delta-\alpha_2), \\
d & = & \cos \alpha_2 \sin 2(\delta-\alpha_2),
\end{eqnarray}

\noindent where the $\alpha$'s are defined by

\begin{equation}
\tan \alpha_{\rm m} = m \, \frac{g_L\,e\,B}{m_{\rm e}\,c} = \frac{m}{2}
	\, \frac{B}{B_{\rm Han}},
\end{equation}

\noindent for $g_L$ the L\'{a}nde factor, and $m=1$ or 2.  The case of
non-magnetic scattering corresponds to $\alpha_{\rm m} = 0$, for
which the scattering source functions reduce to the expressions for
pure Thomson (dipole) scattering except for $E_1$.  In the strong or
``saturated'' limit of the Hanle effect, $\alpha_{\rm m} \approx \pi/2$,
and so $a=b=c=d=0$.  

Finally, it is worth noting the analytic expressions
for $J$ and $K$ in the current approximations.  Assuming a star
of uniform brightness, the Eddington moments are given by

\begin{equation}
J = \frac{1}{4\pi}\,\int \, I_*\,d\Omega = W(r) I_*,
\end{equation}

\noindent and

\begin{equation}
K = \frac{1}{4\pi}\,\int \, I_*\,\mu^2\,d\Omega =
W(r)\, I_* \,
\left(\frac{ 1 + \mu_{\rm c}+ \mu_{\rm c}^2}{3}\right).
\end{equation}

\noindent The factor $\half (3K-J)$ is the familiar finite star depolarization
factor of Cassinelli \etal\ (1987).  From the preceding two expressions,
this geometric factor is given by

\begin{equation}
\half (3K-J) = W(r)\, I_* \times \frac{1}{2}\,\mu_{\rm c}\,\left( 1+
        \mu_{\rm c}\right).
\end{equation}

\noindent At large radius the factor scales like $r^{-2}$, which is
the point source approximation.  However, right at the photosphere, the
it drops to zero.  This results because the symmetry of the radiation
field over half the sky as seen by a scattering particle yields no net
polarization of the scattered light.  On the other hand, limb darkening
prevents the factor from entirely vanishing at the photospheric level.
Although we will ignore limb-darkening in our models, the effect is
relevant for the use of the Hanle effect in studies of the sun that
probe scattering geometries around the solar photospheric level (e.g.,
see Stenflo 1994).

\subsection{The ``Single Scattering'' Approximation for Polarized
Emission Lines}

It is well-known that multiple scattering of linearly polarized light
generally leads to a depolarization.  The polarized intensity from most
media of astrophysical interest is small, and so one can approximate the
radiative transfer as predominantly unpolarized.  The depolarization that
arises from multiple scattering would suggest that an emergent intensity
beam derives the bulk of its polarization from its last scattering
event. In the ``last scattering'' approximation, radiation propagates
through a medium as a beam of unpolarized radiation until it emerges, with
a sense of polarization appropriate for its last scattering interaction.

For the wind case, it is a ``single scattering'' approximation that is the
picture most relevant for our analysis.  In Sobolev theory the radiation
transport for the line emission is localized to relatively small volumes,
sometimes referred to as ``resonance zones''.  In the terminology of
Monte Carlo radiative transfer, one can envision many photon ``packets''
of stellar light that penetrate into these zones, scatter several times,
and emerge toward an observer.  Owing to the probabilistic nature of
treating the scattering, the scattering history of each packet will
be different.  Although each will be characterized by a polarization,
the polarization position angles will on average be random, thus summing
to small net polarizations. In contrast photon packets will typically
scatter only once inside optically thin resonance zones.  These will
all have a polarization, but more importantly they will on average all
have the same polarization position angle so as to sum constructively.
Consequently, we employ the simplifying assumption that the scattered
light from optically thick ($\tau_S > 1$) resonance zones is completely
unpolarized, whereas light emerging from optically thin regions ($\tau_S <
1$) is treated as if single-scattered, for which we use the Hanle source
functions as previously defined.

\section{RESULTS FOR AXIAL AND DIPOLE MAGNETIC FIELDS}

In this section we present polarized profile results for heuristic
purposes, focussing on two simple magnetic geometries: an axial magnetic
field and a dipole magnetic field.  In each case the underlying wind is
assumed to be spherical.  This latter assumption is not dynamically
consistent with the chosen field topologies, but we adopt the condition
so as to isolate the influence of the Hanle effect for the polarization
of wind emission line profiles.

The viewing perspective is described completely by the inclination angle,
$i$ in the stellar frame.  Since many of the strong resonance lines common
to astrophysics are Li-like doublets, we use $E_1=0.5$ for our
calculations, which is the value appropriate for the shorter wavelength
component for many of these doublets (e.g., Chandrasekhar 1960).

\begin{figure}
\plotone{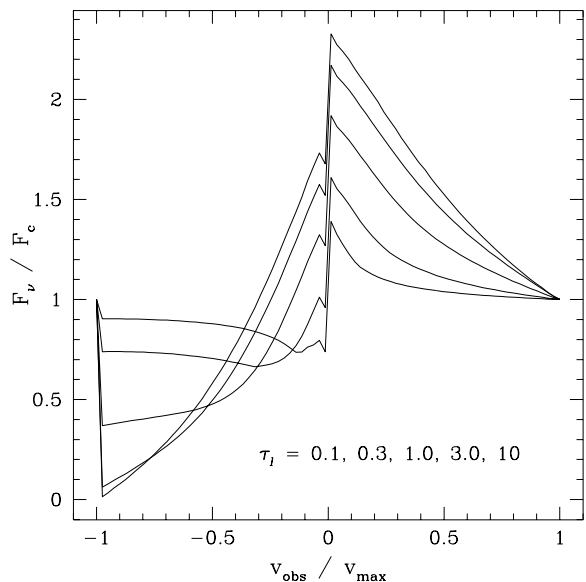}
\caption[]{
A sequence of P~Cygni profiles produced by our code using the escape
probability method with volume emissivity.  The profiles are for optical
depth scales $\tau_l$ as indicated, with thicker lines producing 
deeper blueshifted absorption and stronger redshifted emission.  
The dip that appears just blueward of line center in these profiles
is real.  In each
case the wind velocity law is with $\gamma=1$, and for an ion species
of constant ionization fraction.
\label{fig4}}
\end{figure}

The line strength will be characterized with an optical depth scale given by

\begin{equation}
\tau_l = \frac{\kappa_l\,\rho_0\,\lambda_0}{\vinf/R},
\end{equation}

\noindent and so the Sobolev optical depth can be expressed as

\begin{equation}
\tau_S = \tau_l \, w^{-2}\, x^{-1}\, (1+\mu^2\,\sigma)^{-1},
\end{equation}

\noindent with $w=v/\vinf$ the normalized radial velocity and $x=r/R_*$
the normalized radius.  Some example P~Cygni profiles as calculated
with our numerical code are shown in Figure~\ref{fig4} for a range of
optical depth scales $\tau_l$ and with a $\gamma=1$ velocity law.  All of
our model profile calculations will involve $\gamma=1$, and the case of
$\tau_l=1$ will typically be adopted.  For simplicity we further assume
the ionization of the wind is constant with radius.  The ionization
of different atomic species is generally a function of radius (e.g.,
Drew 1989); however, some ions can be dominant over a substantial range
of radii.  Moreover, although variations in ionization can change the
level of polarized flux, the percent polarization of the profile is not
altered when ionization varies with radius only.  For this reason
we concentrate on the constant ionization case to emphasis the influence
of the Hanle effect for the polarized line emission.

The different magnetic wind models will be parametrized in terms of the
ratio of the surface field strength $B_*$ relative to the Hanle field
$B_{\rm Han}$.  Since the figures will display
polarized line profiles for a range of ratios $B_*/B_{\rm Han}$, one
may view the results in two different fashions.  For a line with a given
Hanle field, different polarized profiles represent how the line
would appear for stars of different surface field strengths.  Conversely,
for a star of a given field strength, the different polarized profiles
represent lines of different Hanle fields from the same wind (i.e.,
assuming that all the lines form over the same spatial region).  It is
useful to scrutinize the results that follow with both perspectives
in mind.

\subsection{The Axial Magnetic Field Case}

For an axial magnetic field, there is a tremendous simplification of the
Hanle effect problem in terms of geometry, because the magnetic field
${\bf B} = B_*\,\hat{z}_*$ is everywhere parallel to the assumed axis of
symmetry for the star with $\vartheta_B=0$ and the field strength is taken
to be constant.  This then is the easiest case
to consider for the Hanle effect.  In fact, the expressions describing
the Stokes $Q$ and $U$ source functions for the light scattered by an
optically thin and uniform ring of matter whose symmetry axis passes
through the origin is analytic.  For a spherically symmetric wind,
such a ring represents the fundamental building block for constructing
theoretical emission line profiles.

For the case at hand -- spherical symmetry in the expansion and density
-- it is the scattered light as integrated around the ring that is
important for our consideration of unresolved sources.  Consider a
ring whose axis is directed toward the observer.  All points on the
ring have the same Doppler shift.  Note that the axis of the ring
is not the same as the symmetry axis describing the magnetic field.
Using equations~(\ref{eq:SII})--(\ref{eq:SUI}) with $\theta_{\rm s}=i$,
$\theta_{\rm i}=\vartheta$, $\delta=-\varphi$ and $i_{\rm s}=0$, the
azimuthally integrated (i.e., in $\alpha$) Stokes $Q$ and $U$ source
functions for axisymmetric rings are found to be

\begin{equation}
S_Q = -\frac{3}{8}\,E_1\,\sin^2 i\,S_0(r)\,\left(1-3\cos^2\chi\right)\,\left[
        (1-3\cos^2 i) + 4\cos^2\alpha_1\,\cos^2 i - \cos^2\alpha_2\,
        (1+\cos^2 i)\right],
\end{equation}

\noindent and

\begin{equation}
S_U = \frac{3}{8}\,E_1\,\cos i\,\sin^2 i\,S_0(r)\,\left(1-3\cos^2\chi\right)
        \,\left(-\sin 2\alpha_1+\frac{1}{2}\sin 2\alpha_2 \right),
\end{equation}

\noindent where $S_0(r)=3K-J$ is a function of radius only, and the Hanle
mixing angles $\alpha_1$ and $\alpha_2$ can vary with location
$(r,\vartheta)$.  Assuming for illustrative purposes that the field
strength varies with radius only, several interesting points can be made
about the Hanle effect in emission lines.

\begin{figure}
\plotone{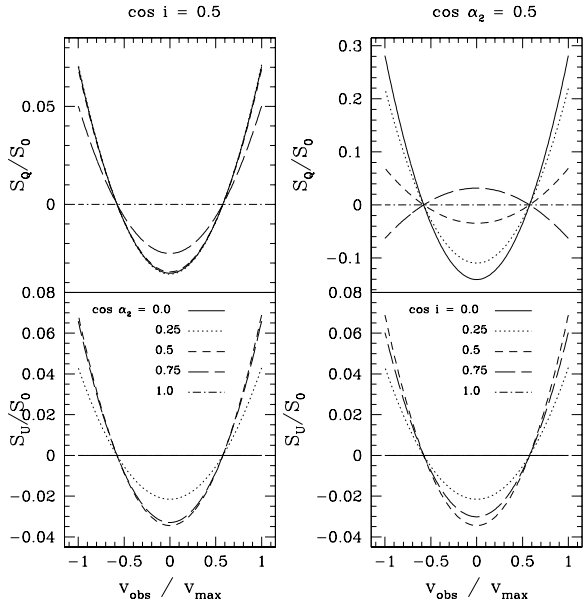}
\caption[]{
Variation of $S_Q$ and $S_U$ across an expanding and optically thin
spherical shell that is threaded by an axial magnetic field of constant
strength.  The Stokes source functions are plotted against observed
Doppler shift, from the front side at ($v_{\rm obs}/v_{\rm max}=-1$)
to the rear side at ($v_{\rm obs}/v_{\rm max}=+1$).  The two panels
at left show the effect of varying the Hanle effect via the parameter
$\cos \alpha_2 = 1.0$, 0.75, 0.5, 0.25, and 0.0 for $B/B_H= 0.0$, 0.9,
1.7, 3.9, and $\infty$.  The panels at right show the effect of varying
the viewing inclination (with the solid line for $i=90^\circ$ and the
dotted short-dashed line for $i=0^\circ$).  Note that the upper right
panel uses a different scale from the other panels.

\label{fig5}}
\end{figure}

Consider an optically thin spherical scattering shell.  Each isovelocity
contour is a ring centered on the observer's viewing axis and
distinguished from other rings by the angle $\chi$.  Figure~\ref{fig5}
shows how the $Q$ and $U$ source functions will vary from ring to ring,
with a constant viewing inclination for the curves at left, and a constant
field strength for the curves at right.

Even though both the ring and field are axisymmetric, note that a
$U$-signal persists. This is because of the field's handedness in
producing the Hanle precession.  Notable points include: (a) Both $Q$
and $U$ vanish for the case $B=0$, since the non-magnetic scattering
polarization vanishes for a spherically symmetric density distribution.
(b) The polarized $Q$ and $U$ source functions both vanish at the
``Van Vleck'' angle, $\cos^2\chi= 1/3$.  (c) The $U$-signal scales with
inclination such that the $U$-polarization vanishes for pole-on and
edge-on viewings.  (d) For spherical expansion the $U$-profile integrates
to zero across the line.  (e) Finally, in the saturated limit of the
Hanle effect, $S_U=0$ and only a $Q$-signal remains.  Indeed, in this
limit the source function $S_Q$ is especially simple with the form

\begin{equation}
S_Q = -\frac{3}{8}\,E_1\,\sin^2 i\,S_0(r)\,\left(1-3\cos^2\chi\right)\,
	\left(1-3\cos^2 i\right).
\end{equation}

\noindent In addition to $S_Q=0$ for the two isovelocity rings
situated at $\cos^2 \chi = 1/3$, the $Q$ source function also vanishes
for viewing inclinations at $\cos^2 i = 1/3$.

\begin{figure}
\plotone{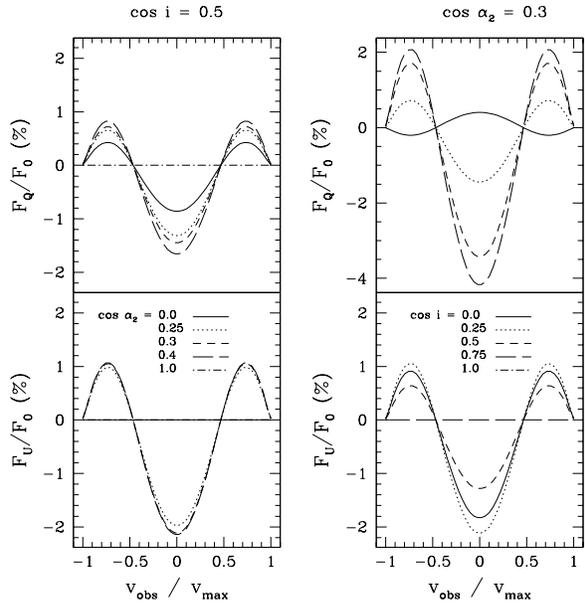}
\caption[]{
Results for the $Q$ and $U$ polarized flux profiles from our
numerical code for an axial magnetic field of constant strength with
$\tau_l=1.0$ and $\gamma=1$.  The sequence of line types are the same
as in Fig.~\ref{fig5}, but the values used for $\cos \alpha_2$ have
been changed slightly so that the profiles may more easily be seen.
The left panels are for $\cos i=0.5$ with different $\alpha_2$ values, and
the right panels are for $\cos \alpha_2=0.3$, but for different values
of $i$.  The vertical scale is different for the upper right panel.
The integrated polarized line flux in both $Q$ and $U$ are generally
non-zero.

\label{fig6}}
\end{figure}

Figure~\ref{fig6} shows the polarized line profile (eq.~[\ref{eq:delLnu}])
for this geometry from our code, assuming a constant field strength and
using the single scattering approximation with $\gamma=1$, $\tau_l=1.0$,
and $E_1=1$.  Displayed are the continuum normalized $Q$ and $U$ Stokes
flux profiles for an axial field and a spherical shell of fixed radius
in percent polarization for the same range of viewing inclinations
(relative to the axial field) as in Figure~\ref{fig5}, but with a
different selection of field strengths to better distinguish between
the model line profiles.  The vertical plot scale is different because
the curves for Figure~\ref{fig5} are source functions, whereas those
in Figure~\ref{fig6} are line fluxes.  Since we explicitly assume that
polarized line emission emerges only from optically thin portions of the
wind, one might expect that the profiles that are evaluated numerically
would resemble those of Figure~\ref{fig5}, at least qualitatively.
And indeed the numerical polarized line profile and the analytic source
function do show some similarity, at line center, but whereas the
analytic derivation predicts relatively strong polarization at the line
wings, the polarized flux drops to zero for the numerical calculations.
In particular, note that the line-integrated $U$ flux does not necessarily
vanish, contrary to expectations.

\begin{figure}
\plotone{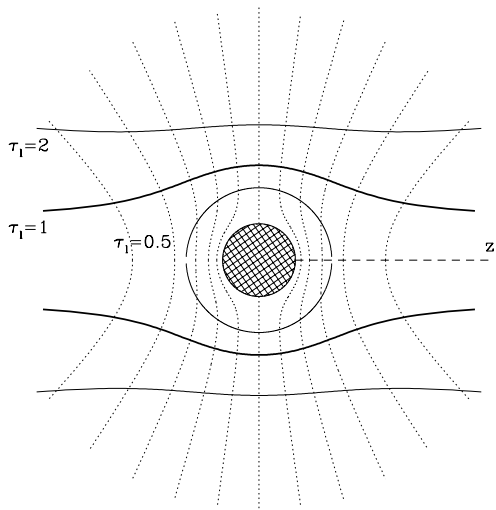}
\caption[]{
Shown as dotted lines are isovelocity zones, with the observer to the
right.  The solid lines are curves demarcating regions of the wind
where the Sobolev optical depth in the line is thick ($\tau_S>1$
interior to the curves) and thin ($\tau_S<1$ exterior to the curves).
Three cases are shown, with $\tau_l =0.5$, 1, and 2.  The $\tau_l=1$
case is shown as a bold line, since it is the value adopted for most of
our line calculations.  In this particular case, some portion of every
isovelocity zone is optically thick in the line.  In the single scattering
approximation, polarized line emission is produced only at
locations where $\tau_S <1$.
\label{fig7}}

\end{figure}

These differences between the source function and line profile can be
understood with the help of Figure~\ref{fig7}, which shows an illustration
with the star at center and dotted lines for isovelocity zones.
The observer is along the $z$-axis located to the right.  The solid
curves are the locus of points with Sobolev optical depth $\tau_S=1$
(assuming a velocity law with $\gamma=1$ and line optical depth scales
$\tau_l$ as indicated).  Interior to sets of  paired solid curves, the
line optical depth is greater than unity, and exterior the line optical
depth is smaller than unity.  The heavy dark curves are for $\tau_l=1$,
which represents the case most typical of our line profile calculations.
In this case the figure indicates that there is no spherical shell that
can be drawn which does not intercept the region of $\tau_S>1$.  Moreover,
the interception, for shells of large radius, occurs preferentially at
those parts of the shell that are primarily fore and aft of the star
with respect to the observer, thus corresponding to Doppler shifts in
the line wings.

Importantly, we discover that there are two new effects that arise from a
consideration of line optical depth effects.  First, although spherical
symmetry is used, the Sobolev optical depth for fixed radius varies
in going from the front side of the shell to the rear side, with some
portions being optically thick and others optically thin.  Second, for
some shells, every point can be optically thick.  For example in the case
$\tau_l=1$, Figure~\ref{fig8} shows that every point inside $r \approx 2.5
R_*$ has $\tau_S > 1$, and so the Hanle effect is effectively ``blind'' to
the circumstellar magnetic fields between the wind base and this radius.

This latter point implies two interesting corollaries.  First, lines
of different strength (as characterized by the different $\tau_l$ values
in the figure) will allow the Hanle effect to probe magnetic fields to
different radial proximities to the wind base.  Second, the Hanle and
Zeeman effects can be used in tandem.  For a star such as $\theta^1$ Ori~C
with a strong multi-kiloGauss surface magnetic field, the Zeeman effect
as applied to photospheric lines yields information about the surface
field distribution and strength.  The Hanle effect as applied to wind
lines will sample the circumstellar magnetic field, which will be weaker
than the surface field.  Together, the line polarizations might be used
to reconstruct how the magnetic field and wind flow affect one another.

In summary, considerations of the Hanle effect in a wind flow with an
axial magnetic field reveals that substantial polarizations of up to 4\%
can result near line center.  The Stokes $Q$ and $U$ profile shapes are
characteristically symmetric, and the polarization position angle does not
vary across the profile.  A point that is significant for observational
studies is that the line-integrated polarized flux does not in general
vanish, implying that a net polarization could be measured even for
poorly resolved line profiles.  The axial field topology is especially
simple, and the derived peak polarizations are to be considered as the
{\em best possible} because a constant vector magnetic field was assumed
everywhere in the wind, whereas in real winds the field strength will
generally diminish with radius and the field orientation will normally
vary from point to point.  The major result of this section is that using
the axial field scenario as a control case, we find that line optical depth
effects do substantially modify the shapes of the polarized profiles
and the portions of the wind that can be probed by the Hanle effect.

\subsection{The Dipole Magnetic Field Case}

Now we apply our theory to extended and radially outflowing envelopes
that are threaded with a dipole magnetic field.  The combination of
a dipole field and a radial wind are not compatible, since either a
strong dipole field will result in wind confinement (Babel \& Montmerle
1997ab; Cassinelli \etal\ 2002) or a strong wind flow will drag out the
magnetic field, distorting the dipole into a more nearly radial geometry
(ud-Doula \& Owocki 2002). Still the dipole case represents a more
complicated field topology than offered by a simple axial field and so
is worth investigating.  Also, this is the topology used in section~3.2
of Paper~II, that had a more simplistic assumption for the effects of
optical depth.  Here we see the influence of optical depth in a better
treatment of the radiative transfer.

It should be mentioned that a dipole magnetic field is purely radial at
the poles, thus by symmetry  yielding no Hanle effect.  In the equatorial
plane, the field is purely axial; everywhere else, the field is a vector
sum of radial and latitudinal components.  At no latitude is there a
toroidal component of the field.  In the next section, the influence of
a toroidal field will be explored in the context of winds from slowly
rotating stars.

\begin{figure}
\plotone{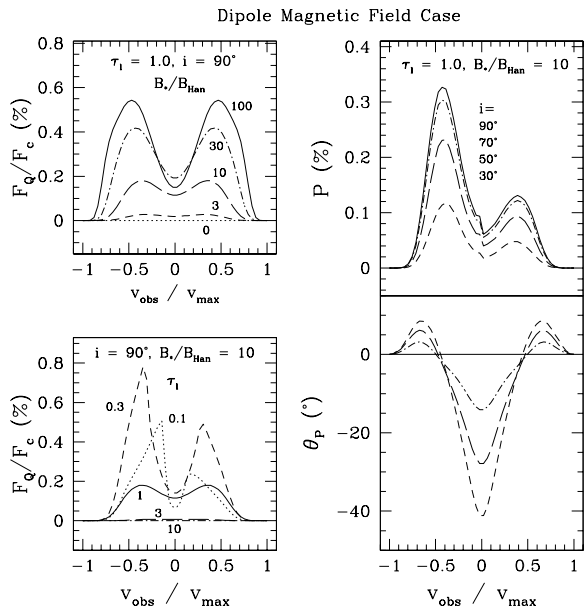}
\caption[]{
The percent polarization of the Hanle effect for a dipole magnetic field
in a spherical stellar wind. (a) Upper left: Shown are polarized Stokes
$Q$ profiles as normalized to the continuum emission {\em outside} the
line frequencies.  For a dipole field as seen from the side in a line
with $\tau_l=1.0$, larger Hanle ratios (as labelled) yield higher line
polarizations.  (b) Lower left: As in (a), except now $\tau_l$ is varied
as indicated for a fixed Hanle ratio.  Note the line asymmetry for the
two lower optical depth cases is a consequence of stellar occultation.
The strongest polarization results for fairly optically thin lines,
that sample stronger fields deeper in the wind.  (c) Upper right:
Plotted is the total polarization $P=\sqrt{F_Q^2+F_U^2}/F_{\rm tot}$
in percent.  The total flux $F_{\rm tot}$ includes the P Cygni line,
which is inherently asymmetric in its profile shape.  Overall, the
effect of viewing inclination is to reduce $P$ for perspectives that are
increasingly pole-on.  (d) Lower right: Same line types as in (c) but now
for the polarization position angle $\theta_P$.  Owing to symmetry, $U=0$
at all points in the line for $i=0^\circ$ and $i=90^\circ$. A left-right
symmetric $U$-profile remains for intermediate viewing inclinations
(which spatially corresponds to fore-aft symmetry about the star).
\label{fig8} }

\end{figure}

Polarized line profiles have been calculated for a range of line
strengths, viewing inclinations, and Hanle ratios, with results displayed
in Figure~\ref{fig8}.  The two panels at left are for edge-on views with
$i=90^\circ$.  In this case Stokes $U$ is identically zero throughout
the line profile owing to the symmetry, and so only Stokes $Q$ is shown
(normalized to the continuum flux $F_{\rm c}$ and plotted in percent).
The profile polarization $P$ and the variation of polarization position
angle $\theta_P$ are shown in the right panels.  The Hanle ratios quoted
in the figure refer to $B_*/B_{\rm Han}$; however, we remind that the
field strength of a dipole diminishes with radius as $B\propto r^{-3}$,
and so a range of radii and field strengths contribute to the observed
line polarization.

Returning to panels at left, the upper one shows that stronger Hanle
ratios lead to lines of higher polarization.  The left-right symmetry
in the polarized profile corresponds to a fore-aft spatial symmetry
with respect to the star.  The panel below illustrates the
effect of line optical depth, with solid for $\tau_l=0.1$, short dash
for $\tau_l=0.3$, long dash for $\tau_l = 1$, dash dot for $\tau_l=3$,
and dotted for $\tau_l=10$.  For small optical depths, the polarized line
emission arises from deep in the wind.  As $\tau_l$ is made to increase,
the region of $\tau_S>1$ moves outward, and the polarization peaks drift
toward larger Doppler shifts in the line.

Initially, increasing $\tau_l$ leads to larger line polarizations, because
there are more scatterers.  For larger optical depths, the region of
$\tau_S >1$ has swelled sufficiently in extent that the overall flux of
polarized line emission begins to drop.  Interestingly, at low values
of $\tau_l$, stellar occultation results in an asymmetric profile of
polarized line flux, with the redshifted emission somewhat suppressed.
For larger $\tau_l$, the profile becomes symmetric.  This can be
understood in terms of Figure~\ref{fig7}.  For large $\tau_l$, all the
points lying directly behind and in front of the star do not contribute
any polarized line emission, because those points all have $\tau_S> 1$,
hence occultation has no influence; at small $\tau_l$ this is not case,
and so occultation makes a difference.

Returning to Figure~\ref{fig8}, the two panels at right show the
effects of viewing inclination, which in the case of a dipole field,
yields both $Q$ and $U$ profiles.  The inclination values are indicated.
The upper plot is the total polarization with $P=\sqrt{F_Q^2+F_U^2}/F_{\rm
tot}$, and the lower plot is for the polarization position angle defined
by $\tan 2\theta_P = F_U/F_Q$.  Note that $F_{\rm tot}$ includes
the redshifted P Cygni line emission and the blueshifted absorption,
making the polarized profile in $P$ asymmetric.  On the whole the line
polarization is seen to drop with viewing inclination, similar to the
$\sin^2 i$ rule of Brown \& McLean (1977) for axisymmetric Thomson
scattering envelopes.

The variation in the polarization position angle is especially
interesting.  Our assumption is that the polarization arises
from the optically thin portions of the wind.  This suggests that
$F_Q$ and $F_U$ are roughly proportional in terms of optical depth
effects.  So the ratio $F_U/F_Q$ that determines the polarization
position angle $\theta_P$ is to first order independent of
optical depth.  Its amplitude is set by the viewing inclination,
and its variation with Doppler shift is governed by the Hanle
effect.  The line types for $\theta_P$ in the lower right plot
of Figure~\ref{fig8} are the same as in the panel above it.
In the absence of a magnetic field, the spherical symmetry ensures
that both $Q$ and $U$ are zero, so that it is the Hanle effect that
produces the line polarization and the position angle variations both.
If the envelope and velocity field were instead axisymmetric, then
position angle rotations could result even without a magnetic field.
However, the Hanle effect would distinguish itself by 
virtue of how the position angle variations differed between
lines of different $A$-values and optical depths.

In comparison to the polarized profiles from spherical winds that
were modelled in Paper~II (Fig.~8 in that work), the profile results
presented here are seen to be qualitatively similar but to differ
somewhat quantitatively.  Under the single scattering approximation,
the polarized profiles have significant polarization near line center,
whereas those of Paper~II were more centrally depressed.  Although the
polarizations in Paper~II were overestimated by about a factor of 2,
the overall asymmetric profile shape in $P$ remains because of the
normalization by the P~Cygni profile.

To sum up, the main lessons to draw from the dipole magnetic field
case are that overall stronger line polarizations result from larger
Hanle ratios, the strongest line polarizations are for lines that are
relatively thin, the polarization is reduced for viewing inclinations
that are further from edge-on, and the influence of the Hanle effect is
betrayed through the observation of position angle rotation effects in
the line profile.  Having established some insight into the Hanle effect
of magnetized winds, and a general sense of the line polarizations
that can result (up to about 0.7\%), we next consider a more
realistic case in the context of winds from slow rotators that drag out
a relatively weak surface magnetic field.

\section{RESULTS FOR A SLOW MAGNETIC ROTATOR}

In this section, in order to define a realistic but tractable magnetic
geometry, we assume that in the case of weak magnetic fields, the star
will have a ``slow magnetic rotator wind''. This means that the field
does not play a role in accelerating the flow to terminal velocity.
We are motivated by the fact that the Sun is a slow magnetic rotator
(Belcher \& MacGregor 1976).  The overall weak solar magnetic field does
not dominate the wind flow, but is believed to have contributed to a
long-term braking of the Sun's rotation (as first analyzed by  Weber \&
Davis 1967).  The field produces some transfer of angular momentum to
the wind, and in the process leads to some rotational distortion of the
flow. For simplicity we also assume that the rotation is slow enough
that the wind velocity field is basically spherical, and that the field
is dragged out into ``streak lines'' by the flow.  Although the velocity
field is treated as a function of radius only, we shall consider both
spherical and axisymmetric density distributions for the line polarization
and the influence of the magnetic fields.

For the underlying field topology, we make use of the WCFields model
(Ignace \etal\ 1998) that describes an initially radial field that is
dragged out by the wind from a rotating star.  The model is based on the
Wind Compression theory of Bjorkman \& Cassinelli (1993).  Expressions
governing the wind flow and magnetic field geometry are summarized
in Appendix~\ref{appB}.  The salient features are that (a) a given
wind model is predominantly determined by the ratio of the equatorial
rotation speed $v_{\rm rot}$ relative to the wind terminal speed \vinf,
(b) the field is initially assumed to be radial at the wind base (i.e.,
a split monopole), (c) the fields are ``weak'' and frozen-in so that
the wind flow geometry determines the magnetic geometry, and (d)
asymptotically, the field becomes dominantly toroidal at equatorial
latitudes while remaining largely radial near the pole.

As described in Appendix~\ref{appB}, we adopt approximate expressions
to describe the flow and magnetic field that can be used for slowly
rotating stars (i.e., $v_{\rm rot} / \vinf \ll 1$).  In this limit
spherical geometry for both the wind velocity and density can be adopted .
As an example, a slow rotator model that can produce interesting line
polarizations is the case $v_{\rm rot} / \vinf =0.08$.  For a terminal
speed of 2000~\kms, this would correspond to a rotation speed of
160~\kms, which is not atypical of observed $v \sin i$ values in O~stars
(e.g., Penny 1996), and amounts to rotating at about 25\% of break-up.
The asymptotic density contrast between the equator and pole, $\rho_{\rm
eq}/\rho_{\rm pole}$, in this case is merely 1.25 (i.e., with $\gamma=1$).
Initially, we shall ignore this density variation in our models but will
return to the issue of how a wind density that deviates from spherical
affects the line polarization.

\subsection{WCFields with a Spherical Wind Density}

\begin{figure}
\plotone{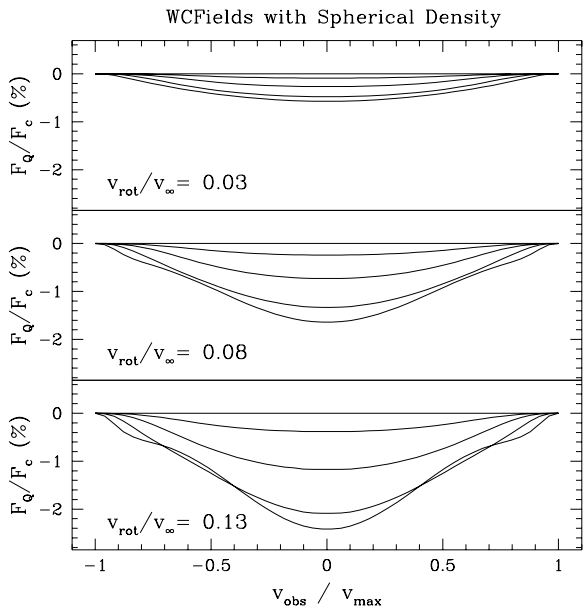}
\caption[]{
Polarized line profiles for slow magnetic rotators at three different
values of $v_{\rm rot}/\vinf$ as indicated.  The magnetic geometries
are determined by the WCFields model.  The wind density and flow velocity
are taken to be spherical, so that the polarization arises only from
the Hanle effect itself.  The different curves in each panel are for
Hanle ratios of $B_*/B_{\rm Han}=0$, 3, 10, 30, and 100, with stronger
polarizations resulting for larger Hanle ratios.  The inclination is
$90^\circ$ and $\tau_l=1$.  A weak $U$ signal exists, but we choose not
to show it here.  \label{fig9}}

\end{figure}

Polarized flux profiles have been computed for the cases $v_{\rm rot} /
\vinf = 0.03$, 0.08, and 0.13, and the results are plotted as percent
polarization in Figure~\ref{fig9}.  Only the $Q$ profiles are plotted.
There is a $U$ profile, but the level of polarization is about an order
of magnitude less than in $Q$, and so we do not show it.  In each panel
the different curves are for different Hanle ratios, with $B_*/B_{\rm
Han}=0$, 3, 10, 30, and 100, such that stronger polarizations result
for larger Hanle ratios.

\begin{figure}
\plotone{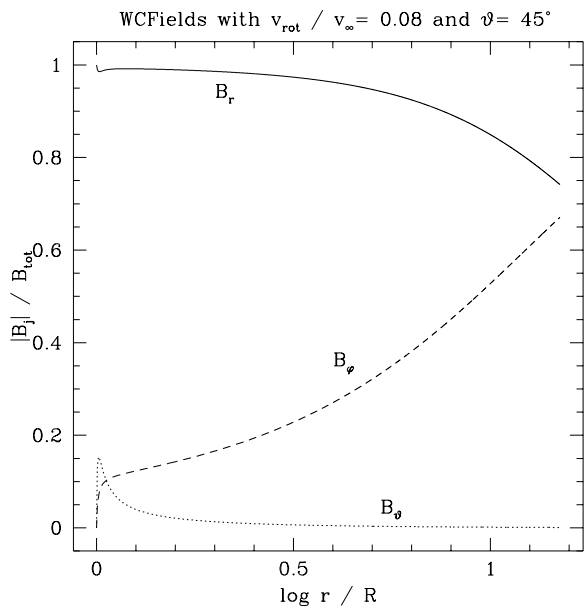}
\caption[]{
Using approximations that apply for slow stellar rotations, this figure
shows the variation of the magnetic field components from the WCFields
model as a function of radius.  The individual components are
normalized to the local total magnetic field strength.  These curves
are for the case $v_{\rm rot}/\vinf=0.08$ at a latitude of
$\vartheta=45^\circ$ so as to roughly maximize the contribution of the
latitudinal component $B_\vartheta$.
\label{fig10}}

\end{figure}

For all of these cases, the polarized profiles are seen to be
single-peaked, in distinction to the dipole field case.  The profiles
also show a net negative polarization, implying a position angle that is
orthogonal to the axis of rotation.  This results from the significant
toroidal magnetic field component, and can be understood as follows.
At the base of the wind, the field is initially a split monopole, with
no toroidal component. Both toroidal and latitudinal components begin to
develop as the flow draws the field out in radius (see Fig.~\ref{fig10}
and App.~\ref{appB}). The radial component of the field decreases
asymptotically as $r^{-2}$, whereas the toroidal component will
decrease as $r^{-1} \sin \vartheta$, and the latitudinal component as
$r^{-3} \sin 2\vartheta$, for $\vartheta$ the stellar co-latitude. At
radii where the Sobolev optical depth is thin, the latitudinal field
component is negligible (although it does account for the small level
of $U$ polarization), and the toroidal component is starting to become
comparable to $B_{\rm r}$.  The field is mainly radial near the poles,
so that in these regions the Hanle effect is minor.  Around the equator
the field is becoming more and more toroidal.

Consider the isovelocity zone corresponding to line center, which
is the plane of the sky as intercepting the center of the star.
Without a magnetic field, the polarized intensity is symmetric about
the line-of-sight to the star. By our convention the polarization is
negative around the polar limb, and positive around the equatorial limb.
The field is radial near the pole, so the polarization of scattered line
radiation does not change in that region.  However near the equator, the
toroidal field is into the plane on one side, out of it on the other.
This leads to the geometry of the classic laboratory Hanle effect,
and in the saturated limit, the scattered light becomes completely
depolarized. Integrating around a circle that is centered on the star and
lies in this plane, it is clear that the net polarization will no longer
be zero, but negative, being dominated by the un-modified polarized flux
from the polar zones.  Similar kinds of arguments can be used to show that
the scattered light from all isovelocity zones will tend to be negative.

Figure~\ref{fig9} shows that the line polarization is stronger for more
rapidly rotating stars.  This just reflects the fact that in WCFields theory,
$B_\varphi \sim v_{\rm rot}/\vinf$, so that for a given location in the
wind, faster rotation implies that the ratio $B_\varphi/B_r$ at a given
radius is larger.  Figure~\ref{fig10} shows the variation of the field
components with radius for a stellar co-latitude of $\vartheta=45^\circ$,
chosen such that the peak value of $B_\vartheta$ will be roughly
maximized.  The field components are plotted as normalized to the total
field strength $B_{\rm tot}=\sqrt{B_r^2+B_\vartheta^2+B_\varphi^2}$.
Since $B_r$ dominates over most of the radii shown, the overall field
strength drops roughly as $r^{-2}$, although somewhat less rapidly at
the larger radii owing to the increasing importance of $B_\varphi$.
As noted previously, at very large radius where the toroidal component
dominates, the field will decrease only inversely with radius, instead of
quadratically.  This figure is for the case of $v_{\rm rot}/\vinf=0.08$;
for faster rotations both the latitudinal and toroidal components would
be relatively larger, and vice versa.

Especially important to realize is that the Hanle effect is a non-linear
effect.  Although the bulk of the scattered light that produces the
polarized line flux comes from radii where $\tau_S$ is of order unity
and somewhat less, Figure~\ref{fig10} indicates that the field is
predominantly radial at these locations which should yield no Hanle
effect.  Thus it is the non-radial components that lead to a Hanle effect
in terms of introducing an {\em asymmetry}, and yet the modification
of the scattering polarization is determined by the {\em total} field
strength, and so the radial field is still relevant.  Consequently, the
Hanle effect is in principle sensitive to the full 3D magnetic field,
in terms of field strength and field geometry.  The main diagnostic
challenge is the fact that the net line polarization at any frequency
in the profile is a strongly convoluted property.  At the very least,
it is evident that larger Hanle ratios lead to greater polarization.
Next we discuss observational prospects and diagnostic approaches.

\begin{figure}
\plotone{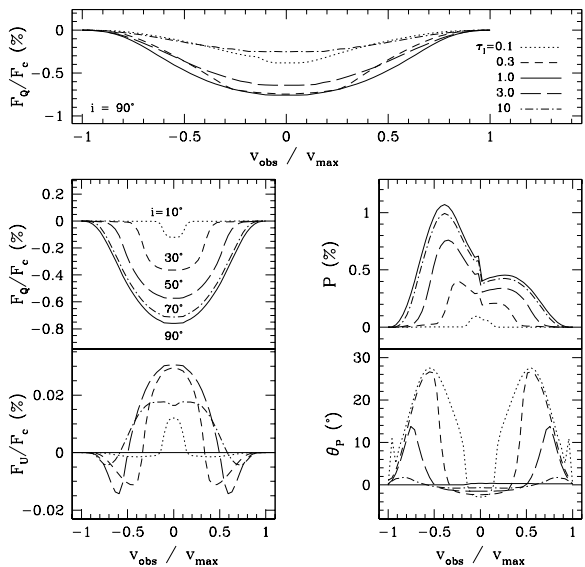}
\caption[]{
A series of polarized line profiles for a WCFields model with $v_{\rm
rot}=0.08\vinf$ to highlight the influence of optical depth and viewing
inclination effects.  The surface field strength is $B_*=10B_{\rm Han}$
for every profile.  The uppermost plot shows the continuum normalized $Q$
flux for $i=90^\circ$ with the line optical depth varied as indicated.
The four panels below show the influence of viewing inclination for
$\tau_l=1.0$: At left are plots of $F_Q/F_{\rm c}$ and $F_U/F_{\rm
c}$ in percent polarization and at right are plots of $P$ (this being
normalized to the full P~Cygni line profile and not just the continuum
level) and $\theta_P$.  The different curves are for the inclination
values shown in the panel for $F_Q$.

\label{fig11}}

\end{figure}

Using the case of $v_{\rm rot}=0.08\vinf$, we have examined the influence
of line optical depth and viewing inclination in greater detail.
Figure~\ref{fig11} shows a series of model line profiles at this fixed
rotation speed as $\tau_l$ and $i$ are varied.  The uppermost panel
shows the effect of line optical depth for the case $i=90^\circ$.
Interestingly, line optical depth had a stronger influence on the line
polarization in the Dipole field case.  Recall that as $\tau_l$ is made
to increase, the region over which the line is optically thick grows in
extent, and this region does not contribute to the line polarization.
The field strength of a dipole decreases as $r^{-3}$, such that for
$\tau_l \gtrsim 1$, the Hanle effect samples the magnetic field where
even at modest radii it is quite small, so that the polarization is
also small.  The WCFields case is different.  The field strength drops
initially as $r^{-2}$ for a split monopole, slowly transitioning to a
predominantly toroidal field that decreases only as $r^{-1}$.  This is
much more gradual than for a Dipole field, so that the line polarization
does not drop significantly until $\tau_l$ becomes exceedingly thick
$\tau_l \gtrsim 3$.  Indeed, the line polarization is larger for a
somewhat thick line at $\tau_l=10$ than for a thin line at $\tau_l =
0.3$, because at low optical depth, most of the optically thin line
radiation comes from the inner the wind, where the field is largely
radial with almost no Hanle effect.

Figure~\ref{fig11} also displays the effect of viewing inclination in
the lower four panels:  the two at left showing polarized flux in $Q$
and $U$, and the two at right showing the total degree of polarization
$P$ and the polarization position angle $\theta_P$.  In each panel the
line type corresponds to the same inclination angle value as indicated
in the panel for $F_Q$.

\subsection{WCFields with an Axisymmetric Wind Density}

A net resonance line scattering polarization occurs even when
there is no Hanle effect (i.e., $B=0$), if the distribution of
scatterers is asymmetric.  The scattering is similar to Thomson
scattering, as previously noted, reduced by the factor $E_1$, but
increased by the much larger cross-section compared to Thomson scattering.

\begin{figure}
\plotone{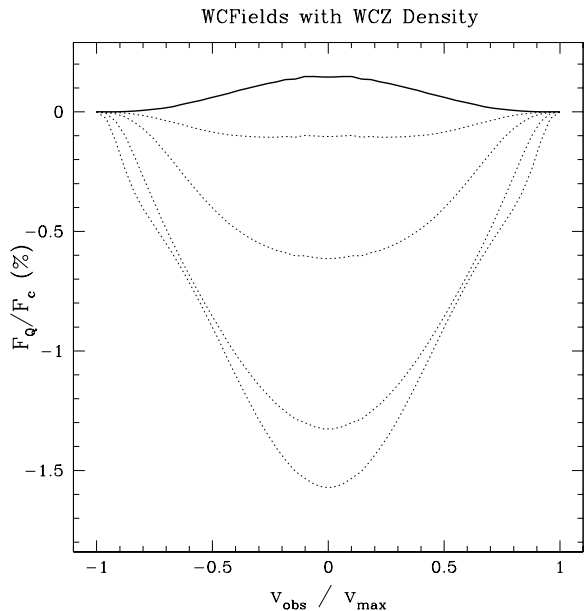}
\caption[]{
The Hanle effect for a WCFields model with $v_{\rm rot}/\vinf=0.08$ but
now with an aspherical wind.  The bold solid line is the line
polarization that results when there is no magnetic field.  The dotted
lines are with the Hanle effect (to be compared with the middle panel
of Fig.~\ref{fig9}.  The view is edge-on and $\tau_l=1$ for all
of the profiles.  For $B=0$, peak polarization
occurs near line center and is positive because there are relatively
more line scatterers in the vicinity of the equator,
than near the poles.
\label{fig12}}

\end{figure}

The previous section ignored this polarization contribution by assuming
the wind density to be spherical.  The motivation was to isolate the
influence of the Hanle effect.  Now relaxing this approximation,
Figure~\ref{fig12} shows polarized line profile shapes for the case
$v_{\rm rot}/\vinf = 0.08$ and $\gamma = 1$, which has an equator to
pole density contrast of 1.25.  The upper dark line is the polarization
from pure resonance scattering with no Hanle effect.  The dotted
curves are the profiles with the Hanle effect, to be compared with
Figure~\ref{fig9} that assumed an underlying spherical wind density.
So the density distribution does introduce a kind of bias for the
line polarization for its interpretation in terms of the Hanle effect.
However, it is not a large effect, and a strategy that targets lines
with different Hanle fields should be able to correct for the influence
of non-spherical density.

\begin{figure}
\plotone{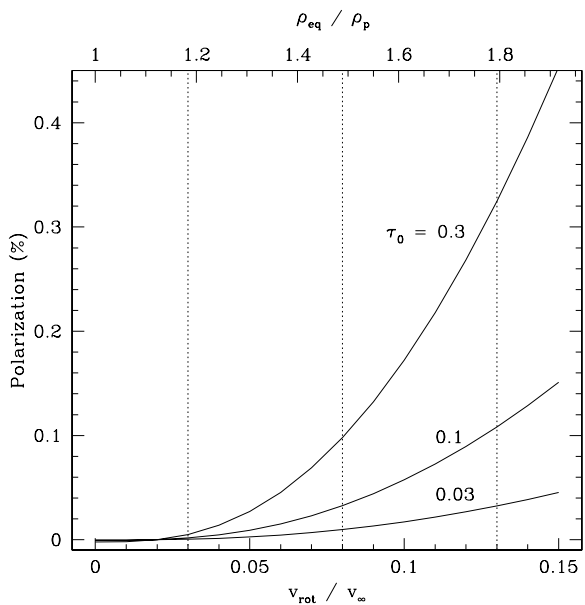}
\caption[]{
Continuum polarizations arising from Thomson scattering of starlight in
the mildly distorted WCZ models.  The abscissa is $v_{\rm rot}/\vinf$
relevant for the models presented in this paper.  Also shown at top is the
corresponding asymptotic equator-to-pole density contrast.  The electron
scattering is assumed to be optically thin, with envelope optical depths
$\tau_0$ as indicated.  These $\tau_0$ values are for an equivalent
spherical envelope.  The continuum polarizations are for an edge-on view,
and will scale as $\sin^2 i$ for other viewing inclinations.  The filled
dots indicate the corresponding models used in the line calculations
for the WCFields magnetic geometry assuming a spherical wind density.

\label{fig13}}

\end{figure}

Allowing for $\rho(r,\mu)$ also implies that scattering polarization by
free electrons in the wind may complicate the interpretation of the
line polarizations.  Figure~\ref{fig13} shows the expected continuum
polarization from mildly distorted stellar winds as viewed edge-on. For
this figure we are using a sequence of Wind-Compressed Zone (WCZ)
models (Ignace, Cassinelli, \& Bjorkman 1996) in conjunction with the
expressions of Brown \& McLean (1977) for the polarization from optically
thin axisymmetric envelopes. The results also include the finite star
depolarization factor (Cassinelli \etal\ 1987). The lower axis is the
ratio $v_{\rm rot}/\vinf$, and the upper axis is the asymptotic value
of the equator-to-pole density contrast.  Electron scattering depth of
$\tau_0=0.03$, 0.1, and 0.3 for an equivalent spherical wind were used
as reference models.  The three points indicate the rotation values
used for the Hanle effect models presented in the previous section.
Relatively few O~stars are known to be intrinsically polarized at the
0.1\% level, but typically $\tau_0 < 0.1$, so this is not inconsistent
with the small distortions implied in our models (McDavid 2000).

In any case continuum polarization of any kind, including interstellar
polarization, is not a significant complication to our analysis: It turns
out that for the strong Li-like doublets commonly observed in the UV and
FUV band, the red component has $E_1=0$, meaning that the line scatters
isotropically, making no contribution to the line polarization and thus
not susceptible to the Hanle effect. Any polarized flux observed in the
red line of the doublet must therefore arise from continuum processes,
and since the doublet components are close together in frequency, the
polarization of the blue component which can show a Hanle effect may be
straightforwardly corrected for the continuum polarization.

\subsection{Field Topology}

\begin{figure}
\plotone{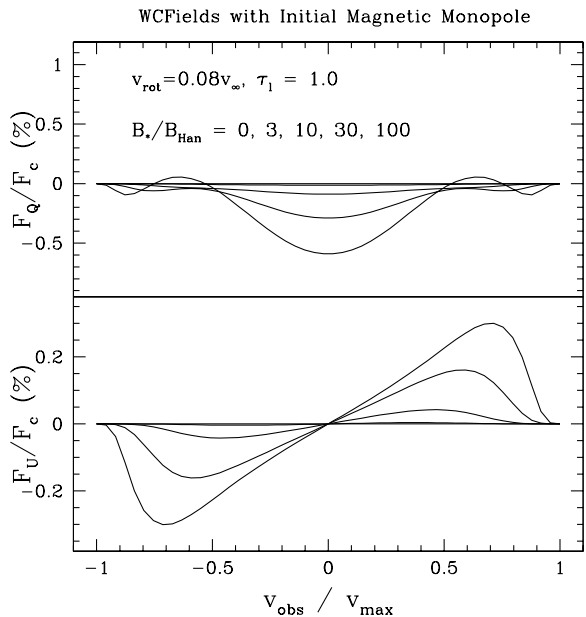}
\caption[]{
Polarized line profiles in $Q$ and $U$ for a WCFields model, with
$v_{\rm rot}/\vinf=0.08$ and a spherical density, but now with an
initial magnetic field at the wind base that is a magnetic monopole.
Compared to the basal split monopole configuration used in Fig.~\ref{fig9},
the $Q$ polarization has dropped, and a significant and antisymmetric
$U$ profile is now evident.  Consequently, the $U$ profile, when the
polarization measurement axis is aligned with the source symmetry axis,
is sensitive to the symmetry of the circumstellar magnetic geometry.
\label{fig14}}

\end{figure}

The Hanle effect is sensitive both to the field strength and the magnetic
geometry.  This sensitivity can be made especially graphic.  We previously
noted that for the WCFields models, there is a small $U$ polarization.
In Figure~\ref{fig14}, we plot the polarized profiles for a model with
$v_{\rm rot} / \vinf = 0.08$, $i=90^\circ$, and a spherical wind, but
now for a magnetic field that is initially outward radial at the wind
base (i.e., a magnetic monopole -- unphysical, but used here only to
make a point).  The result is that the $Q$ profiles are much reduced
in scale, and a fairly substantial anti-symmetric $U$ profile results.
The models are identical in all respects to the corresponding models of
Figure~\ref{fig9}, except that the radial field in the lower hemisphere is
outwardly directed instead of inwardly directed.  Thus we find that it is
still the case that larger Hanle ratios yield stronger line polarizations,
but moreover, that the $U$ profile is sensitive to the overall symmetry
(top-bottom and left-right) of the global magnetic geometry.

In practice, the symmetry axis of the magnetic field (if one exists)
will not be known {\it a priori}.  The Stokes $Q$ and $U$ line profiles
for a source are measured according to observer-defined axes.  It may
be possible to identify symmetry in the magnetic field by rotating the
observed polarizations using Mueller matrices to look for symmetry in
the polarized profiles.  Generally, a constraint on the symmetry of the
circumstellar magnetic field would be estimated as part of the profile
modeling.

\section{DISCUSSION}	\label{sec:disc}

This contribution has focussed on the Hanle effect in resonance scattering
lines common to hot star winds.  We have stressed the fact that the Hanle
effect is non-linear, in that it is the non-radial field components that
give rise a change in the line polarization, yet at the same time, it is
the total field strength (including the radial component) that governs
the amount of the change.  Consequently, the Hanle effect can be used
to infer the full 3D magnetic geometry in the wind.  For a given set of
scattering lines, stronger fields will lead to higher line polarizations.
Field topology also has an influence, as evidenced by the double-peaked
polarized profiles for a dipole magnetic field, versus the single-peaked
profiles for slow magnetic rotators.

The Far Ultraviolet Spectro-Polarimeter (FUSP) is a sounding rocket
payload with the goal of making the first spectro-polarimetric
measurements in the wavelength range of 1050--1500~\AA\ for several
stellar targets (Nordsieck 2003).  With a 50~cm primary, the instrument
will have a spectral resolution of 0.65~\AA, corresponding to a
resolving power of $\lambda/\Delta \lambda = 1800$ at a wavelength of
1170~\AA.  This in turn corresponds to a velocity resolution of about
170~km~s$^{-1}$, which is sufficient to resolve wind-broadened lines
with typical full widths of a few thousand km~s$^{-1}$.  Targeting the
star $\zeta$~Ori, which was observed in the FUV with Copernicus (Snow \&
Morton 1976), it is anticipated that FUSP will produce the first detection
of the Hanle effect in a star other than the Sun.

To correctly interpret the polarization of scattering lines that will
be measurable with FUSP, care must be taken to identify all sources of
polarized radiation.  For example, interstellar polarization is endemic
to all polarimetric studies of distant objects, and there are standard
techniques for its correction, based on wavelength dependence and
lack of variability.

Extracting the influence of a non-spherical wind geometry for the line
scattering polarization can be approached in two different ways.  (a)
If the interstellar polarization can be determined, the polarization
outside line frequencies due to Thomson scattering can be used to
model the underlying wind geometry.  However, this is difficult in part
because it may not be clear that the line forms in the same region that
gives rise to the electron scattering polarization.  (b) Alternatively,
since ${\bf B}({\bf r})$ is fixed for any given star, one can use a
multi-line approach to infer the polarization arising from an aspherical
distribution of scatterers.  Different lines have different Hanle field
values, so that lines with large Hanle values can be used to estimate
the non-magnetic contribution.  Once again, care must be taken since
different lines may not form in the same spatial locations; however,
the P~Cygni profiles themselves could be modelled to obtain information
about the ionization distribution in radius and latitude.

The study of the Hanle effect for circumstellar environs and the
development of radiative transport techniques is ongoing, especially in
terms of relaxing the single scattering approximation for the line
polarization.  In this regard one can use expressions in Jeffery
(1990) for ``Sobolev-P'' theory to show that the single scattering
approximation captures the flavor of the polarimetric behavior of
scattering lines, but that in fact there is non-zero polarization from
line scattering even at rather high Sobolev optical depths of a few,
and even at optical depths of just a few tenths, the line polarization
is not truly described by single scattering.  We anticipate using Monte
Carlo radiation transport techniques in the future to model the line
polarization more accurately. The proven versatility of these techniques
will be useful in handling more realistic wind models.

Another effect ignored here needs to be explored. While our analysis
applies directly to resonance transitions in ions with no nuclear spin
(e.g., \ion{O}{6}, \ion{Si}{4}, \ion{C}{4}, \ion{Mg}{2}, and \ion{Ca}{2}),
in other Lithium-like ions such as \ion{P}{5}, \ion{N}{5}, \ion{Na}{1} and
\ion{K}{1}, the presence of nuclear spin breaks the degeneracy of the spin
states, and this alters the value of the polarizability $E_1$ making it
sensitive to optical pumping.  In fact in the presence of optical pumping,
the magnetic field has another effect on the line polarization through
``magnetic re-alignment'' (Nordsieck 2001), or in Solar Physics parlance,
the ``second Hanle effect''.  We are investigating the regime in which
the Hanle effect and the magnetic re-alignment physics both apply.

\acknowledgements We thank Jon Bjorkman for useful discussions regarding
line polarization effects.  We are especially grateful to the referee,
Marianne Faurobert-Scholl, for several helpful comments.  Support for
this research comes from a grant from the National Science Foundation
(AST-0098597).

\appendix

\section{DETERMINATION OF ANGULAR QUANTITIES}	\label{appA}

The Hanle effect is complicated by the fact that three coordinate systems
are involved:  star, observer, and the local magnetic field orientation.
A standard approach for solving for the requisite angles is to employ
spherical trigonometry.  More challenging are the interior angles of
the spherical triangles in part because of quadrant ambiguity and in
part because of potential divide-by-zero problems.  An equivalent and
more straight-forward approach is to employ vector relations, which we
describe here.

The givens include the stellar and observer axes and the magnetic
field geometry.  These immediately provide the scattering direction,
and vector magnetic field at any point.  The stellar axes also allow
one to specify the direction of the incident radiation throughout the
scattering medium.

For the various spherical triangles of Fig.~\ref{fig3}, the ``arc
relations'' are given by dot products, with

\begin{eqnarray}
\cos \vartheta_{\rm s} & = & \hat{z}\cdot\hat{z_*} , \\
\cos \vartheta_{\rm i} & = & \hat{r}\cdot\hat{z_*} , \\
\cos \chi & = & \hat{r}\cdot\hat{z} , \\
\cos \vartheta_B & = & \hat{z_*}\cdot\hat{B} , \\
\mus & = & \hat{r}\cdot\hat{B} , \\
\mui & = & \hat{z}\cdot\hat{B},
\end{eqnarray}

\noindent where $\hat{z}=\hat{s}$.

\begin{figure}
\plotone{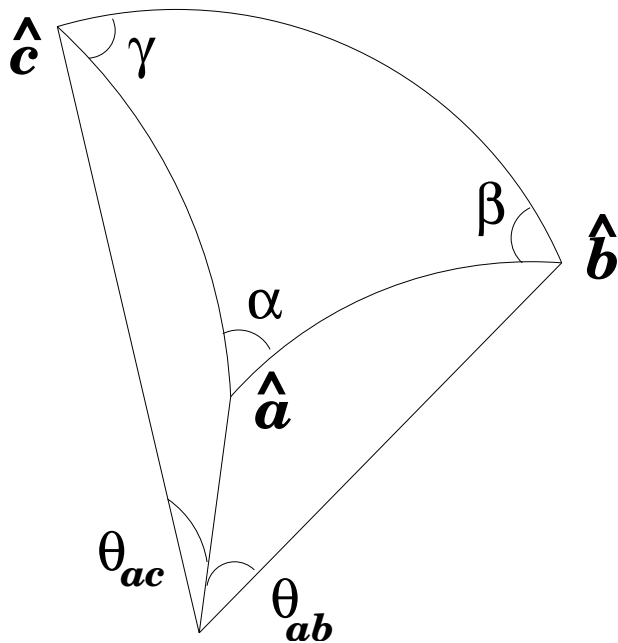}
\caption[]{
Specification of spherical triangle unit vectors, spherical polar angles,
and interior angles used in relating the evaluation of the angles by
vector product relations.  \label{fig:app1}}

\end{figure}

Consider the spherical triangle of
Fig.~\ref{fig:app1}, drawn from unit vectors $\hat{a}$, $\hat{b}$, and
$\hat{c}$.  Let the interior angles be $\alpha$, $\beta$, and $\gamma$,
respectively.  Moreover, let the arcs be specified by $\theta_{\rm ac}$,
$\theta_{\rm ab}$, and $\theta_{\rm bc}$.  Then one can determine the
cosine and sine of an interior angle by deriving the unit tangent vectors
at the vertices of interest.

For example to determine the angle $\alpha$, one defines the unit tangent vectors
$\hat{t}_{\rm ab}$ and $\hat{t}_{\rm ac}$ at the vertex of $\hat{a}$ in
the directions of $\hat{b}$ and $\hat{c}$.  Then one has that

\begin{equation}
\cos \alpha = \hat{t}_{\rm ab} \cdot \hat{t}_{\rm ac},
\end{equation}

\noindent and

\begin{equation}
\sin  \alpha = \hat{a} \cdot (\hat{t}_{\rm ac} \times \hat{t}_{\rm ab} ).
\end{equation}

\noindent where the unit vectors are given by

\begin{equation}
\hat{t}_{\rm ab} = \frac{1}{\sin \theta_{\rm ab}}\, \left[ (\hat{a}
	\times \hat{b})\times \hat{a} \right],
\end{equation}

\noindent and

\begin{equation}
\hat{t}_{\rm ac} = \frac{1}{\sin \theta_{\rm ac}}\, \left[ \hat{a}
	\times (\hat{c}\times\hat{a} ) \right]
\end{equation}

For the Hanle effect with incident unpolarized radiation, the
two interior angles of relevance are $i_{\rm s}$ and $\delta$.
For the first of these, and working through the vector relations,
one obtains

\begin{eqnarray}
\cos i_{\rm s} & = & (\xxs\,\sin\vartheta_{\rm s})^{-1}\,
	\left[ \hat{B}\cdot\hat{z_*}-(\hat{z}\cdot\hat{z_*})
	(\hat{B}\cdot\hat{z})\right], \\
\sin i_{\rm s} & = & (\xxs\,\sin\vartheta_{\rm s})^{-1}\,
	\left[ \hat{z}\cdot(\hat{B}\times\hat{z_*})\right].
\end{eqnarray}

\noindent and for the second angle, one has

\begin{eqnarray}
\cos\delta & = & (\xxs\xxi)^{-1}\,
        \left[ \hat{r}\cdot\hat{z}-(\hat{z}\cdot\hat{B})
        (\hat{B}\cdot\hat{r})\right], \\
\sin \delta & = & -(\xxs\,\xxi)^{-1}\,
        \left[ \hat{B}\cdot(\hat{z}\times\hat{r})\right].
\end{eqnarray}

\section{WCFIELDS FOR SLOW ROTATORS}	\label{appB}

In Wind Compression theory (Bjorkman \& Cassinelli 1993; Ignace \etal\
1996), the distortion of the wind flow is described kinematically.
Ignoring pressure gradient terms, and assuming only radial forces, one
can derive an expression for streamline flow.  The general properties of
wind compression have been confirmed with hydrodynamic simulations by
Owocki, Cranmer, \& Blondin (1994), although the effects appear to be
inhibited when non-radial accelerations that arise in the line-driving
of hot star winds are included (Owocki, Cranmer, \& Gayley 1996).

However, retaining the basic elements of Wind Compression theory,
Ignace \etal\ (1998) extended the method to allow for ``weak''
magnetic fields.  The key assumptions are that the magnetic field
is frozen-in and dominated by the hydrodynamic flow.
Consequently, the known flow geometry determines the magnetic field
topology.

Referring to Ignace \etal\ (1998) for the derivation, the key expression
that determines the vector magnetic field throughout the flow is

\begin{equation}
{\bf B} = B_*\,\left(\frac{R_*^2}{r^2}\right)\,\left(\frac{d\mu}{d\mu_0}
	\right)^{-1}\,\frac{{\bf V}}{v_{\rm r}},
\end{equation}

\noindent where $B_*$ is the surface field strength, $d\mu/d\mu_0$
is the ``compression factor'' that describes how neighboring streamlines
evolve throughout the wind flow, ${\bf V}$ is the vector velocity in
the co-rotating frame, and $v_{\rm r}$ is the radial velocity.
All of the velocity components and the compression factor are knowns.
The scalar strength of the field is 

\begin{equation}
B = B_*\,\left(\frac{R_*^2}{r^2}\right)\,\left(\frac{d\mu}{d\mu_0}
	\right)^{-1}\,\left\{1+\frac{v_{\rm rot}^2}{v_{\rm r}^2}\,
	\left[ \frac{\sin^2\vartheta_0\,\cos^2\vartheta_0\,\sin^2\phi '}
	{\sin^2\vartheta} + \left(\frac{R_*}{r}\sin\vartheta_0
	-\frac{r}{R_*}\sin\vartheta\right)^2\right]\right\}^{1/2},
\end{equation}

\noindent where $v_{\rm rot}$ is the equatorial rotation speed of the
star, and $\phi '$ is a ``deflection'' angle that determines the
streamline flow and is described in Bjorkman \& Cassinell (1993).  The
zero subscript indicates a value at the base of the wind.  In the
upper hemisphere, streamlines
move toward the equator with $\vartheta > \vartheta_0$ for $r>R_*$.
In the lower hemisphere, streamlines also move toward the equator, now
with $\vartheta < \vartheta_0$.
The field is taken to be radial at the wind
base (split monopole), but tends to toroidal configuration at large
radius, with $B\propto r^{-1}\,\sin\vartheta$, like the magnetic field
that is dragged out in the solar wind.

The vector orientation of the magnetic field at any point is given by
the angles $\vartheta_B$ and $\varphi_B$ defined by the relations

\begin{equation}
\cos \vartheta_B = \hat{z_*}\cdot \hat{B} = \frac{B_{\rm r}}{B}\,
	\cos \vartheta - \frac{B_\vartheta}{B}\,\sin \vartheta,
\end{equation}

\noindent and

\begin{equation}
\tan \varphi_B = \frac{B_{y_*}}{B_{x_*}} = \frac{B_{\rm r}\sin\vartheta
	\sin \varphi+B_\varphi\cos\varphi+B_\vartheta\cos\vartheta\sin
	\varphi}{B_{\rm r}\sin\vartheta\cos
	\varphi-B_\varphi\sin\varphi+B_\vartheta\cos\vartheta\cos
	\varphi}.
\end{equation}

So far, these expressions are valid for all rotation speeds, for which the
streamlines are not equator-crossing.  The assumption of slow rotation
with only mild distortions of the wind flow from spherical implies the
following simplifications (Ignace 1996; Bjorkman \& Ignace in prep).

First, the terminal speed depends only weakly on the initial latitude
of a streamline and may thus be assumed a constant.

Second, the streamline parameter becomes $\phi'\approx
\phi_{\rm eq}' (r)\sin\theta_0$, a separable function of radius and initial stellar
colatitude, with

\begin{equation}
\phi_{\rm eq}'(r) = \frac{v_{\rm rot}}{\gamma\,v_0}\,\left(\frac{v_0}{\vinf-v_0}
	\right)^{1/\gamma}\, B_Y(1/\gamma,1-1/\gamma),
\end{equation}

\noindent where $B_Y$ is the incomplete Beta function, with $Y=1-v_0/
v_{\rm r}$.

Third, the radial and latitudinal dependence of
the compression factor can be fit with the following form,

\begin{equation}
\left(\frac{d\mu}{d\mu_0}\right)^{-1} \approx (1+\phi_{\rm eq}'^2)^{-1} +
	\left[ (\cos \phi_{\rm eq}')^{-1}-(1+\phi_{\rm eq}'^2)^{-1} \right] \sin^q\vartheta,
\end{equation}

\noindent where the exponent factor $q$ is

\begin{equation}
q(r) = 3 \tan \phi_{\rm eq}'(r).
\end{equation}

\begin{figure}
\plotone{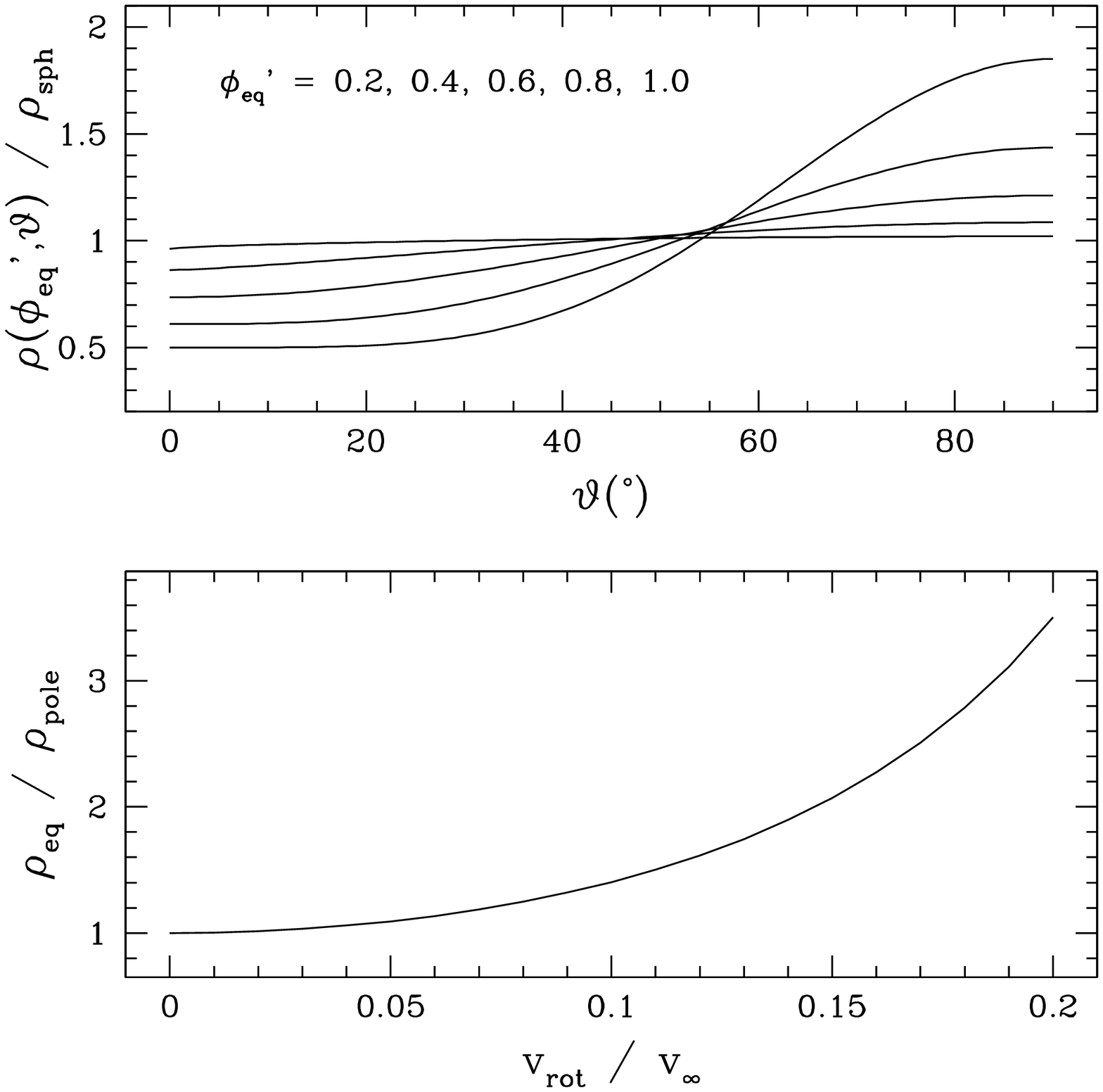}
\caption[]{
The wind geometry of WCZ models using our approximations for slowly
rotating stars.  The upper panel shows the variation of the normalized
wind density with co-latitude $\vartheta$ at different values of
$\phi_{\rm eq}'$.  Below is the asymptotic wind density contrast between
the equator and the pole as a function of the stellar rotation normalized
to the wind terminal speed.

\label{fig:app2}}

\end{figure}

\noindent Consequently, the wind density as a function of radius and
latitude is described by

\begin{equation}
\rho(r,\mu) = \rho_{\rm sph}\,\left(\frac{d\mu}{d\mu_0}\right)^{-1},
\end{equation}

\noindent where $\rho_{\rm sph}= \rho_0\,x^{-2}\,w^{-1}$.  The upper
panel of Fig.~\ref{fig:app2} shows the variation of the wind density
with co-latitude $\vartheta$ at different values of $\phi_{\rm eq}'$.
The lower panel displays the asymptotic equator-to-pole density contrast
$\rho_{\rm eq} / \rho_{\rm pole}$ as a function of the ratio $v_{\rm
rot}/\vinf$.

Finally, with regard to the magnetic field, it is adequate to assume that
$\vartheta_0 \approx \vartheta$, which yields for the total field
strength

\begin{equation}
B = B_*\,\left(\frac{R_*^2}{r^2}\right)\,\left(\frac{d\mu}{d\mu_0}
	\right)^{-1}\,\left\{1+\frac{v_{\rm rot}^2}{v_{\rm r}^2}\,
	\left[ \cos^2\vartheta\,\sin^2\phi '
	+ \sin^2\vartheta\,\left(\frac{R_*}{r}
	-\frac{r}{R_*}\right)^2\right]\right\}^{1/2}.
\end{equation}

In the special case that $\gamma=1$, the function $\phi_{\rm eq}'(r)$ takes
on an especially simple form, with

\begin{equation}
\phi_{\rm eq}'(r) = \frac{v_{\rm rot}}{\vinf}\,\ln \left(\frac{v_{\rm r}}
	{v_0}\right).
	\label{eq:g1}
\end{equation}

\noindent Since $\gamma=1$ is the wind velocity used throughout our
model line profile calculations, Eq.~(\ref{eq:g1}) used in conjunction
with the preceding expressions define the flow geometry and the
magnetic field throughout the wind.

\twocolumn

\begin{references}

\reference{} Babel, J., \& Montmerle, T.  1997a, A\&A, 323, 121
\reference{} Babel, J., \& Montmerle, T.  1997b, ApJ, 485, L29
\reference{} Belcher, J.~W., \& MacGregor, K.B.  1976, ApJ, 210, 498
\reference{} Bjorkman, J.~E., \& Cassinelli, J.~P.  1993, ApJ, 409, 429
\reference{} Brown, J.~C., \& McLean, I.~S.  1977, A\&A, 57, 141
\reference{} Brown, J.~C., Carlaw, V.~A., \& Cassinelli, J.~P.
	1989, ApJ, 344, 341
\reference{} Cassinelli, J.~P., Nordsieck, K.~H., \& Murison, M.~A.
	1987, ApJ, 317, 290
\reference{} Cassinelli, J.~P., \& Maheswaran, M.  1992, ApJ, 386, 695
\reference{} Cassinelli, J.~P., Nordsieck, K.~H., \& Ignace, R.  2001,
	in Magnetic Fields Across the Hertzsprung-Russell Diagram,
	ASP Conf.~248, ed.\ G.~Mathys, S.~Solanki, \& D.~Wickramasinghe
	(San Francisco: ASP), 409
\reference{} Cassinelli, J.~P., Brown, J.~C., Maheswaran, M., Miller,
	N.~A., \& Telfer, D.~C.  2002, ApJ, 577, 951
\reference{} Castor, J.~I.  1970, MNRAS, 149, 111
\reference{} Chandrasekhar, S.  1960, Radiative Transfer (New York: Dover)
\reference{} Donati, J.-F., Wade, G.~A., Babel, J., Henrichs, H.~F.,
	de Jong, J.~A., \etal  2001, MNRAS, 326, 1265
\reference{} Donati, J.-F., Babel, J., Harries, T.~J., Howarth, I.~D.,
	Petit, P., \etal  2002, MNRAS, 334, 374
\reference{} Drew, J.  1989, ApJS, 71, 267
\reference{} Hamilton, D.~R.  1947, ApJ, 106, 457
\reference{} Hanle, W.  1924, Z.~Phys., 30, 93
\reference{} Henrichs, H.~F.  2002, to appear in Magnetic Fields
	in O, B and A Stars:  Origin and Relation to Pulsation, Rotation
	and Mass Loss, (eds.) Balona, Henrichs, and Medupe
\reference{} Ignace, R.  1996, Ph.D.~Thesis, Univ.~Wisconsin
\reference{} Ignace, Cassinelli, J.~P., \& Bjorkman, J.~E.  1996, ApJ,
	459, 671
\reference{} Ignace, R., Nordsieck, K.~H., \& Cassinelli, J.~P. 1997, ApJ,
	486, 550 (Paper I)
\reference{} Ignace, R., Cassinelli, J.~P., \& Bjorkman, J.~E.  1998, ApJ,
	505, 910
\reference{} Ignace, R., Cassinelli, J.~P., \& Nordsieck, K.~H.  1999,
	ApJ, 520, 335 (Paper II)
\reference{} Ignace, R. 2001, ApJ, 547, 393 (Paper III)
\reference{} Ignace, R., \& Gayley, K.~G.  2003, MNRAS, 341, 179
\reference{} Jeffery, D.~J.  1990, ApJ, 352, 267
\reference{} Lin, H., Penn, M.~J., \& Kuhn, J.~R.  1998, ApJm, 493, 978
\reference{} McDavid, D.  2000, AJ, 119, 352
\reference{} Mihalas, D.  1978, Stellar Atmospheres (San Francisco: Freeman)
\reference{} Mitchell, A.~C.~G., \& Zemansky, M.~W.  1934, Resonance Radiation
	and Excited Atoms (Cambridge: University Press)
\reference{} Moruzzi, G., \& Strumia, F., eds.  1991, The Hanle Effect and
	Level-Crossing Spectroscopy (New York: Plenum Press)
\reference{} Neiner, C.  2002, Ph.D Thesis (U.~of Amsterdam)
\reference{} Nordsieck, K.~H.  2001, in Magnetic Fields Across the
	Hertzsprung-Russell Diagram, ASP Conf.~248, ed.\ G.~Mathys,
	S.~Solanki, \& D.~Wickramasinghe (San Francisco: ASP), 607
\reference{} Nordsieck, K.~H., Jaehnig, K.~P., Burgh, E.~B., Kobulnicky,
	H.~A., Percival, J.~W., \etal  2003, SPIE, 4843, 170
\reference{} Owocki, S.~P., 0ranmer, S.~R., \& Blondin, J.~M.  1994, ApJ,
	424, 887
\reference{} Owocki, S.~P., Cranmer, S.~R., \& Gayley, K.~G.  1996, ApJ,
	472, L115
\reference{} Penny, L.~R.  1996, ApJ, 463, 737
\reference{} Rybicki, G.~B., \& Hummer, D.~G.  1978, ApJ, 219, 654
\reference{} Rybicki, G.~B., \& Hummer, D.~G.  1983, ApJ, 274, 380
\reference{} Snow, T.~P., Jr., \& Morton, D.~C.  1976, ApJS, 32, 429
\reference{} Sobolev, V.~V.  1960, Moving Envelopes of Stars
	(Cambridge:  Harvard University Press)
\reference{} Stenflo, J.~O.  1994, Solar Magnetic Fields (Dordrecht: Kluwer)
\reference{} Stenflo, J.~O.  1998, A\&A, 338, 301
\reference{} ud-Doula, A., \& Owocki, S.~P.  2002, ApJ, 576, 413
\reference{} Weber, D.~J., \& Davis, L., Jr.  1967, ApJ, 148, 217

\end{references}
\end{document}